\documentclass[twocolumn,tighten]{aastex62}
\pdfoutput=1 
\usepackage{amsmath,amstext}
\usepackage[T1]{fontenc}
\usepackage{apjfonts}
\usepackage[figure,figure*]{hypcap}
\input{hyperlink-year-only-natbib-patch}

\newcommand*{\ihod}{iHOD}
\newcommand*{\https}[1]{\href{https://#1}{#1}}
\newcommand*{\http}[1]{\href{http://#1}{#1}}
\newcommand*{\msun}{\text{M}_\odot}
\newcommand*{\weburl}{\https{portal.nersc.gov/project/lsst/descqa/v1/?run=2017-11-27}}

\newcommand{\rev}[1]{#1}

\shorttitle{DESCQA}
\shortauthors{Mao et al.\ (LSST~DESC)}

\begin{document}
\title{DESCQA: An Automated Validation Framework for Synthetic Sky Catalogs}

\author[0000-0002-1200-0820]{Yao-Yuan~Mao}
\affiliation{Department of Physics and Astronomy and the Pittsburgh Particle Physics, Astrophysics and Cosmology Center (PITT PACC), University of Pittsburgh, Pittsburgh, PA 15260, USA}
\author{Eve~Kovacs}
\affiliation{Argonne National Laboratory, Lemont, IL 60439, USA}
\author[0000-0003-1468-8232]{Katrin~Heitmann}
\affiliation{Argonne National Laboratory, Lemont, IL 60439, USA}
\author{Thomas~D.~Uram}
\affiliation{Argonne National Laboratory, Lemont, IL 60439, USA}
\author[0000-0001-5501-6008]{Andrew~J.~Benson}
\affiliation{Carnegie Observatories, Pasadena, CA 91101, USA}
\author{Duncan~Campbell}
\affiliation{McWilliams Center for Cosmology and Department of Physics, Carnegie Mellon University, Pittsburgh, PA 15213, USA}
\author{Sof\'ia~A.~Cora}
\affiliation{Instituto de Astrof\'isica de La Plata (CCT La Plata, CONICET, UNLP), Paseo del Bosque s/n, B1900FWA, La Plata, Argentina}
\affiliation{Facultad de Ciencias Astron\'omicas y Geof\'isicas, Universidad Nacional de La Plata, Observatorio Astron\'omico, Paseo del Bosque, B1900FWA La Plata, Argentina}
\author{Joseph~DeRose}
\affiliation{Kavli Institute for Particle Astrophysics and Cosmology \& Department of Physics, Stanford University, Stanford, CA 94305, USA}
\author{Tiziana~Di~Matteo}
\affiliation{McWilliams Center for Cosmology and Department of Physics, Carnegie Mellon University, Pittsburgh, PA 15213, USA}
\author[0000-0002-7832-0771]{Salman~Habib}
\affiliation{Argonne National Laboratory, Lemont, IL 60439, USA}
\author[0000-0003-2219-6852]{Andrew~P.~Hearin}
\affiliation{Argonne National Laboratory, Lemont, IL 60439, USA}
\author[0000-0002-6825-5283]{J.~Bryce~Kalmbach}
\affiliation{Department of Physics, University of Washington, Seattle, WA 98105, USA}
\author[0000-0002-4410-7868]{K.~Simon~Krughoff}
\affiliation{Large Synoptic Survey Telescope, 950 N Cherry Avenue, Tucson, AZ 85719, USA}
\author[0000-0001-7956-0542]{Fran\c{c}ois~Lanusse}
\affiliation{McWilliams Center for Cosmology and Department of Physics, Carnegie Mellon University, Pittsburgh, PA 15213, USA}
\author{Zarija~Luki\'c}
\affiliation{Lawrence Berkeley National Laboratory, Berkeley, CA 94720, USA}
\author[0000-0003-2271-1527]{Rachel~Mandelbaum}
\affiliation{McWilliams Center for Cosmology and Department of Physics, Carnegie Mellon University, Pittsburgh, PA 15213, USA}
\author[0000-0001-8684-2222]{Jeffrey~A.~Newman}
\affiliation{Department of Physics and Astronomy and the Pittsburgh Particle Physics, Astrophysics and Cosmology Center (PITT PACC), University of Pittsburgh, Pittsburgh, PA 15260, USA}
\author[0000-0001-9850-9419]{Nelson~Padilla}
\affiliation{Instituto de Astrof\'isica, Pontificia Universidad Cat\'olica de Chile, Av.\ Vicu\~na Mackenna 4860, Santiago, Chile}
\author[0000-0002-4637-2868]{Enrique~Paillas}
\affiliation{Instituto de Astrof\'isica, Pontificia Universidad Cat\'olica de Chile, Av.\ Vicu\~na Mackenna 4860, Santiago, Chile}
\author[0000-0003-2265-5262]{Adrian~Pope}
\affiliation{Argonne National Laboratory, Lemont, IL 60439, USA}
\author[0000-0002-5294-0630]{Paul~M.~Ricker}
\affiliation{Department of Astronomy, University of Illinois, Urbana, IL 61801, USA}
\author[0000-0001-5035-4913]{Andr\'es~N.~Ruiz}
\affiliation{Instituto de Astronom\'ia Te\'orica y Experimental (CONICET-UNC) and Observatorio Astron\'omico (UNC), Laprida 854, X5000BGR, C\'ordoba, Argentina}
\author{Ananth~Tenneti}
\affiliation{McWilliams Center for Cosmology and Department of Physics, Carnegie Mellon University, Pittsburgh, PA 15213, USA}
\author[0000-0002-4998-7606]{Cristian~A.~Vega-Mart\'inez}
\affiliation{Instituto de Astrof\'isica de La Plata (CCT La Plata, CONICET, UNLP), Paseo del Bosque s/n, B1900FWA, La Plata, Argentina}
\author[0000-0003-2229-011X]{Risa~H.~Wechsler}
\affiliation{Kavli Institute for Particle Astrophysics and Cosmology \& Department of Physics, Stanford University, Stanford, CA 94305, USA}
\affiliation{SLAC National Accelerator Laboratory, Menlo Park, CA, 94025, USA}
\author{Rongpu~Zhou}
\affiliation{Department of Physics and Astronomy and the Pittsburgh Particle Physics, Astrophysics and Cosmology Center (PITT PACC), University of Pittsburgh, Pittsburgh, PA 15260, USA}
\author[0000-0001-6966-6925]{Ying~Zu}
\affiliation{Department of Astronomy, Shanghai Jiao Tong University, 955 Jianchuan Road, Shanghai 200240, People's Republic of China}
\affiliation{Center for Cosmology and AstroParticle Physics (CCAPP), Ohio State University, Columbus, OH 43210, USA}
\collaboration{(The LSST Dark Energy Science Collaboration)}

\begin{abstract}
The use of high-quality simulated sky catalogs is essential for the success of cosmological surveys. The catalogs have diverse applications, such as investigating signatures of fundamental physics in cosmological observables, understanding the effect of systematic uncertainties on measured signals and testing mitigation strategies for reducing these uncertainties, aiding analysis pipeline development and testing, and survey strategy optimization. The list of applications is growing with improvements in the quality of the catalogs and the details that they can provide. Given the importance of simulated catalogs, it is critical to provide rigorous validation protocols that enable both catalog providers and users to assess the quality of the catalogs in a straightforward and comprehensive way. For this purpose, we have developed the DESCQA framework for the Large Synoptic Survey Telescope Dark Energy Science Collaboration as well as for the broader community. The goal of DESCQA is to enable the inspection, validation, and comparison of an inhomogeneous set of synthetic catalogs via the provision of a common interface within an automated framework. In this paper, we present the design concept and first implementation of DESCQA. In order to establish and demonstrate its full functionality we use a set of interim catalogs and validation tests. We highlight several important aspects, both technical and scientific, that require thoughtful consideration when designing a validation framework, including validation metrics and how these metrics impose requirements on the synthetic sky catalogs.
\end{abstract}


\accepted{January 8, 2018}
\submitjournal{the Astrophysical Journal Supplement}

\section{Introduction}
\label{sec:intro}

The Large Synoptic Survey Telescope (LSST) will conduct the most comprehensive optical imaging survey of the sky to date, yielding a wealth of data for astronomical and cosmological studies. LSST data will offer many exciting scientific opportunities, including the creation of very detailed maps of the distribution of galaxies, studies of transient objects in new regimes, investigations of the inner and outer solar system, observations of stellar populations in the Milky Way and nearby galaxies, and studies of the structure of the Milky Way disk and halo and other objects in the Local Volume.
A broad description of the LSST scientific goals is provided in the LSST Science Book~\citep{2009arXiv0912.0201L}.
 
Several science collaborations have been formed in order to prepare for the arrival of the rich and complex LSST data set. One of these collaborations, the LSST Dark Energy Science Collaboration (LSST~DESC), focuses on the major dark energy investigations that can be carried out with LSST, including weak- and strong-lensing measurements, baryon acoustic oscillations, large-scale structure (LSS) measurements, supernova distances, and galaxy cluster abundance. An overview of LSST DESC goals is provided in the White Paper authored by the \cite{2012arXiv1211.0310L}; a detailed Science Road Map can be found at the Collaboration Web site\footnote{\http{lsstdesc.org}}.

Science opportunities relevant to LSST DESC will pose many new analysis challenges on different fronts, including controlling systematic errors, extracting subtle signals from large data sets, combining different dark energy probes, cross-correlations with other observations, etc. The best tools for these tasks must extract the science of interest while simultaneously enabling control of systematic contaminants, whether observational or modeling induced. Before the data set arrives, robust synthetic sky catalogs that are validated against a range of observational data are essential to the development of the required analysis tools. 

{The comprehensive and systematic validation of synthetic sky catalogs presents major challenges. Observational data sets used for validation must be carefully curated and frequently updated with the best available measurements. Tests comparing observations with synthetic data sets have to be designed to address the wide range of tasks for which the catalogs will be used, e.g., tests of photometric pipelines, extraction of cosmological parameters, mass estimates for clusters,
etc. The list is essentially as long as the set of analysis
tasks to be covered by the survey. For each of these tasks, a set of requirements, such as accurate clustering statistics, best possible match to observed colors, detailed galaxy properties, or results over a range of different redshift epochs, needs to be defined and implemented as part of the validation tests.
The synthetic catalogs will be revised, enhanced, and improved over time; a controlled and easy-to-use mechanism to expose new synthetic catalogs to a full battery of observational tests is essential to validating the catalogs properly. In addition, for the users of the catalogs, it is very desirable to have a convenient method to check the catalog quality for their specific needs.}
{In order to provide an environment that can address all of these challenges in a streamlined way, we present DESCQA, a validation framework that has been configured to compare and validate multiple different synthetic sky catalogs in an automated fashion.} 

{There are a number of requirements that must be met in designing a comprehensive and flexible framework intended to assess the performance of multiple synthetic sky catalogs.}
The synthetic catalogs that the system has to handle will have a range of very different characteristics and features depending on how the catalogs are constructed and their ultimate purpose. There are many different ways to build synthetic sky catalogs. Because of the large survey volumes required for cosmological studies, most current methods are based on gravity-only simulations rather than on the significantly more expensive hydrodynamic methods. A range of approaches is used to assign galaxies to dark matter halos in post-processing.
These include halo occupation distribution modeling \citep[HOD;][]{1997MNRAS.286..795K,1998ApJ...494....1J,2000MNRAS.318.1144P,2000MNRAS.318..203S,2002ApJ...575..587B,2002MNRAS.329..246B,2005ApJ...633..791Z, mandelbaum06, zm15a}, subhalo abundance matching \citep[SHAM;][]{Vale2004,Conroy2006,Behroozi10,2010ApJ...710..903M,2010MNRAS.403.1072W,Reddick2013}, and semi-analytic modeling \citep[SAM;][]{1991ApJ...379...52W,1993MNRAS.264..201K,1994MNRAS.271..781C,1999MNRAS.310.1087S,2000MNRAS.311..793B,2003ApJ...599...38B,2006RPPh...69.3101B,2010PhR...495...33B}.
The choice of method used depends on the base simulation (e.g., resolution, available information with regard to time evolution) and the observational data sets that the catalog is expected to match. All known methods are, at best, only partially predictive, and each individual choice has its own advantages and disadvantages, such as resolution requirements, predictive power, ease of implementation, time to run, etc.
In addition, different science cases impose different requirements on the catalogs. Roughly speaking, resolving smaller physical scales increases modeling difficulty, while including broader classes of galaxies adds to the complexity. The galaxy properties required will also influence the choice of method employed.

Given the current uncertainties in galaxy modeling, it is not possible to address the full range of science issues with only one catalog construction method (or a single base simulation). Instead, catalog providers choose the methods that are best suited to address specific questions (or classes of such questions) of interest.
{The heterogeneity among the catalogs manifests itself in both the implementation details, such as file formats, units, and quantity labels, and scientific details, such as the choice of halo-finder algorithms, mass definitions, and filter-band definitions. This heterogeneity presents a significant barrier for users who wish to use several of these catalogs. The framework therefore needs to be capable of ingesting a wide range of synthetic sky catalogs that have very different intrinsic characteristics.

The framework's success hinges on a set of well-thought-out validation tests that can act seamlessly on the catalogs. The tests, as well as the criteria for how well a catalog performs in a given test, will be provided by domain experts (e.g., members of analysis working groups). They will have the best understanding of the requirements and often also of the validation data. The difficulty of adding a new test to the framework therefore has to be minimal so that most contributions can be made without significant assistance from the framework developers. In addition, the execution of the test, once it is implemented, needs to be carried out automatically on all catalogs that provide the necessary information for the test. Tests and catalogs will be improved over time and new ones will be added. The framework needs to provide straightforward methods to accommodate these updates and additions.

Finally, since the validation tests will be run on a wide range of synthetic sky catalogs, it is very desirable to have a convenient method to check which catalogs meet the specific needs for certain tasks. The results must be presented in a way that gives the catalog users an easy-to-use interface to peruse the different tests and catalogs. At the same time, the interface should provide a useful and easily interpretable overview. For this reason, sets of summary statistics and simple assessment criteria of the catalog quality need to be provided.

To summarize, a validation framework used by a survey collaboration such as LSST~DESC should be able to (1) process a wide range of heterogeneous catalogs, (2) automate the validation tests, (3) provide straightforward methods to update and add catalogs and tests, and (4) provide easy access to catalog feature summaries and quality assessments for catalog users. }

{DESCQA addresses the above requirements and
provides interfaces to access synthetic catalogs and to run a set of pre-specified validation tests.} The results of the tests are graphically displayed and evaluated via well-defined error metrics.
DESCQA is accessed via a Web-based portal, which is currently set up at the National Energy Research Scientific Computing Center (NERSC). The portal itself can, in principle, be set up anywhere, but collocating the portal with the storage and analysis resources is convenient.

{In order to demonstrate the full functionality of the framework and to have a sufficiently complex environment for testing and development, it is vital to use realistic, scientifically interesting synthetic sky catalogs and validation tests.} 
{LSST~DESC is still in the process of defining the requirements for each of its analyses and producing new synthetic sky catalogs (including comprehensive full-coverage light-cone catalogs). Therefore, we have chosen to implement a set of interim validation tests and requirements to assist us in the development of the current version of DESCQA{; we present these tests as our case studies of the DESCQA framework. Also, since one major requirement of this framework is to process a wide range of heterogeneous catalogs, we also select a set of interim catalogs that cover the major synthetic methods to be used in our case studies.}
}

{The interim catalogs and tests presented in this work fulfill functions beyond the provision of demonstration examples.
Although these catalogs and tests are not the final versions that will be used for LSST DESC science, their realistically complicated features provide a unique opportunity to delve into the conceptual challenges of building a validation framework.
These challenges originate from the different choices made by the creators of the catalogs and tests, such as the definitions of physical quantities. These intrinsic differences cannot be easily homogenized by the framework; however, the framework can highlight them for the scientists who use the framework.
Working with this set of interim catalogs and tests, we have identified several such conceptual challenges.}
Furthermore, our implementation of the validation framework also provides a concrete platform for publicly and quantitatively {\em defining} the requirements for a particular scientific analysis.

The paper is organized as follows. {We first describe the design and implementation of  the DESCQA framework} in \autoref{sec:descqa}. We explain in detail our method for adding synthetic catalogs to the framework and show how the method enables the automated testing and validation of these catalogs. We also discuss how the Web interface helps the user to navigate the catalogs and validation tests. 
{Then, in \autoref{sec:validation}, we present our case studies of five interim validation tests to demonstrate the features of the framework. The description of the different methods employed to build a range of interim synthetic sky catalogs can be found in \autoref{sec:catalogs}.}
We conclude in \autoref{sec:conclusion} with a summary and discussion of DESCQA and future development paths.


\section{DESCQA Framework}
\label{sec:descqa}

In this section, we describe the DESCQA framework, a unified environment for comparing different synthetic catalogs and data sets using a number of validation tests.

{The DESCQA framework is based on DESQA, which was originally developed for validating catalogs for the Dark Energy Survey. DESQA, in turn, originated from the FlashTest framework, which was developed for standard regression testing (software testing) of the Flash code. Since regression testing is a considerably simpler task than the validation of catalogs, we had to make multiple changes to the framework to accommodate the design goals discussed in \autoref{sec:intro}. 
The basic structure common to all of the variants of the framework is a set of scripts that execute the tests and a Web interface that displays the results. For DESCQA, although much of the original framework has been revised or replaced, the use of the Python programming language is retained \rev{(but revised to be compatible with Python 3)}, along with some portion of the original Web interface, and several of the original concepts used in FlashTest.}

\begin{figure}[tb!]
\centering \includegraphics[width=0.85\columnwidth]{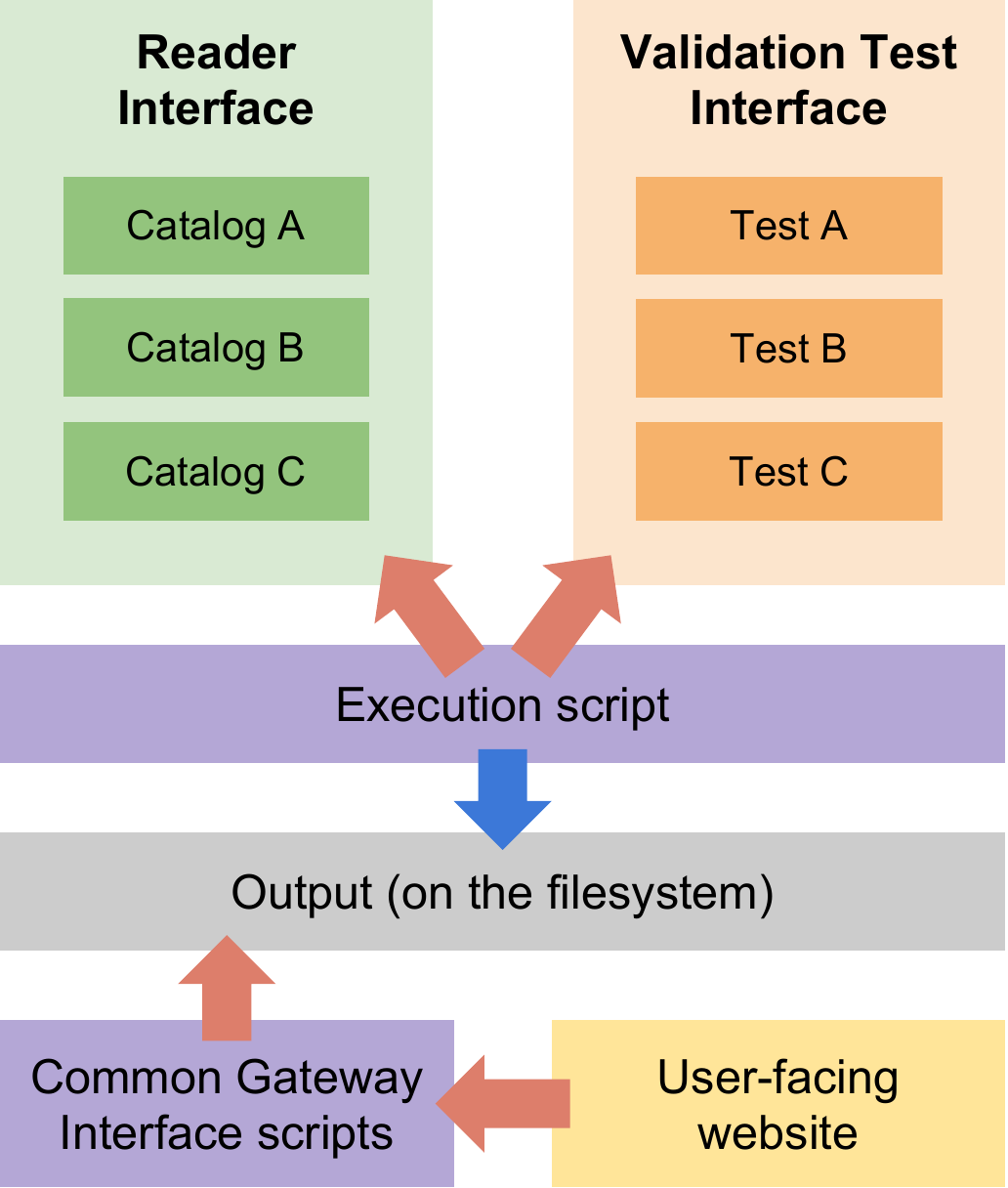}
\caption{\label{fig:framework}%
Illustration of the DESCQA framework.
The two purple boxes are the main drivers of this framework. 
The execution script, which is designed to be triggered periodically, accesses (red arrows) the available catalogs and tests through the reader interface and the validation test interface respectively, and then writes out (blue arrow) the results (including scores and figures) on the filesystem. 
The Common Gateway Interface scripts, which are triggered when a user accesses the Web interface, read the test results that are available on the filesystem, and present them as browsable Web pages.}
\end{figure}

\autoref{fig:framework} presents the organization of the framework, which possesses four main components: (1) the reader interface, (2) the validation test interface, (3) the automated execution process, and (4) the Web interface. 
Together, they enable an expandable, automatable, and trackable validation process.
The code of the DESCQA framework is publicly available in a GitHub repository\footnote{\label{fn:github}\https{github.com/LSSTDESC/descqa}}.
\rev{A frozen version of the code can be found in \citet{descqa-v2.0.0-0.4.5}.}

\subsection{Design Guidelines}
\label{sec:design}

{In designing the framework for DESCQA, our priorities were to provide a unified environment for a set of validation tests that examine and compare different synthetic catalogs, while ensuring that new catalogs and validation tests can be added or updated with minimal effort required from the framework developers, the test developers, and the catalog providers. At the same time, we also want to ensure that the validation results generated by the framework and delivered to the user are easy to understand and compare.}

{The above set of considerations all aim at minimizing the overhead in meeting the requirements imposed by the framework. To help achieve this goal, we separate the framework into three major components, which are as independent as possible. These components are detailed in the following sections: the reader interface, the validation test interface, and the Web interface.}
{In fact, the only requirement for the catalog providers and test developers is to conform to these common application programming interfaces (APIs) for accessing the catalogs (with the reader interface) and for executing the validation tests (with the validation test interface). There is \emph{no} formal requirement regarding the underlying implementation.
The flexibility of Python enables providers and users to implement their readers and validation tests using methods ranging from reading existing files to running an external executable, as long as they provide a Python interface that is consistent with the API specification.}

{In practice, most catalog providers (test developers) need similar high-level functionality in their catalog readers (validation tests). From the point of view of code development, it is desirable to reduce code duplication in order to maintain consistency and reduce human errors. We have designed base classes for both catalog readers and validation tests. Catalog providers and test developers write subclasses that inherit from the base classes, but overwrite the core functions with their own, if needed.

All of the above features reflect our design philosophy of providing an efficient, flexible, automated framework that is capable of including a diverse set of synthetic catalogs and validation tests with the fewest possible requirements imposed on the contributors and code developers. 
}

\subsection{Reader Interface}
\label{sec:reader}

Given the heterogeneity among different synthetic catalogs, it is impractical to access all the different catalogs directly in their native formats within the framework. The standardized reader interface solves this problem by associating with each catalog a corresponding \emph{reader} that loads the catalog in its native format, provides metadata and a list of galaxy properties available in that catalog, and processes any necessary unit \rev{or definition} conversions. The catalog provider implements the corresponding \emph{reader} that conforms to the specification of the standardized reader interface. In this fashion, all of the catalogs can be stored in their native formats, and no change to the catalog's generation pipeline is required.  
\rev{Similarly, the reader can also be used to fix minor errors in the native catalogs (e.g., incorrect definitions or units) during the conversion between native quantities and user-facing quantities, thereby reducing the number of catalog files while still propagating updates to the users.}

{We implemented a base class that contains some basic catalog process methods. To implement a new reader for a specific catalog, one would first subclass this base class, and then supply or overwrite the methods (e.g., data loading routine) to accommodate the catalog under consideration.
Since we expect that different versions of a specific kind of catalog would be accessed using the same code, we allow the same reader to be used with different configuration parameters that specify, for example, the paths of the catalog files and versions. 
These configuration parameters are passed as arguments when the subclass is initialized.
\rev{The reader can also check the catalog's version against an online catalog repository and warn the user if the catalog in use is out of date.}}

{Although some catalog variations such as units and quantity labels can be homogenized by the reader interface, others such as mass definitions and cosmology cannot. We do \emph{not} ask the catalog providers to conform to a specific list of standards, but instead ask that they specify their choices as metadata available in the readers. Similarly, we do \emph{not} require the catalog providers to include all quantities that are needed for all validation tests. During execution, a catalog that does not have some requested quantities for a specific test will be skipped and noted.}

In this particular study, the reader interface is used to enforce consistent units across different catalogs. We use comoving Mpc (not scaled to $h=1$) for all distance units, physical km\,s$^{-1}$ for all velocity units, and $\msun$ (not scaled to $h=1$) for all masses (including stellar mass and halo mass).

\rev{Note that the reader interface itself can actually do more than serving data to the DESCQA framework. It can, in fact, be used as a standalone catalog data server or as a converter to convert catalogs from their native format into a database with common schema. We package the reader interface as a standalone Python module,\footnote{\https{github.com/LSSTDESC/gcr-catalogs}} which allows people to access the homogenized synthetic galaxy catalogs conveniently outside the validation framework. Under this new structure, the DESCQA framework itself becomes a user of the reader interface.}

\subsection{Validation Test Interface}
\label{sec:qa}

An important part of our framework is the quality assurance (QA) implementation, which allows test developers to design validation tests and provides a convenient interface for users to assess the quality of the synthetic catalogs. {In DESCQA, a \emph{validation test} is carried out to establish whether a synthetic catalog meets some particular requirements that have been set by the test developer.} {The validation test consists of two parts. First, the catalog is checked to see if it provides the quantities required for the specific test. Next, the catalog is tested to see if it can reproduce relevant observational data over a specified range at the required accuracy.}

We have designed a standardized validation test interface which is similar in concept to the reader interface. 
{We implemented a base class with abstract methods for the validation test interface. Each individual test is a subclass of this base class and contains the non-abstract methods to conduct the test. 
The test interface also separates the configuration from the code that carries out the actual computation to allow convenient changes to the specific settings for each test.}

Each test uses specified catalog quantities (already wrapped by the reader interface) as input, carries out necessary calculations, creates figures and reduced data, and finally returns a summary statistic. As mentioned in \autoref{sec:design}, if a catalog does not provide all of the required quantities for a particular test, the test will automatically skip the catalog and proceed with the remaining catalogs.

Each test must provide a summary statistic (score) for each catalog on which it runs and also provide a score threshold to determine if a catalog ``passes'' the test. The score and the passing threshold are both up to the test developer to set in the most useful way for that particular test.
The notion of ``passing'' and ``failing'' a test is intended to give the user a quick method to inspect the summarized results using the Web interface, as we detail below. The notion is not to judge the quality of a catalog, as each catalog has its own features. Furthermore, many users are only interested in a subset of validation tests, so a catalog does not need to pass every test in order to be scientifically useful. 

{In addition to the score, the framework allows the validation tests to generate figures, tables, and other products. These supplementary products are saved on the filesystem so that they can be accessed by the web-interface component of the framework as described below. Many users also find it helpful to have a summary figure that displays the relevant statistics for all available catalogs. Although the validation test interface does not formally require all tests to generate plots, we do provide a convenient plotting module to produce basic summary figures which test developers can utilize. Alternatively, developers may supply their own plotting modules.  We will show examples of these summary figures that are generated by our common plotting module in \autoref{sec:validation} for each of our currently implemented tests. }

{We should note that currently all validation tests implemented in this framework are for demonstration purposes. Although this set of tests represents the major tests that are relevant to LSST~DESC science, the choice of summary statistics and the passing criteria presented here are preliminary and introduced only as interim values. In future, test developers and catalog users will set more realistic criteria by which to evaluate the catalogs according to the LSST~DESC science goals.}

\subsection{Automated Execution and Web Interface}

Since our design separates the configurations and the actual reader and test code (which are implemented as classes), a master execution script is required to run the tests. 
For both catalogs and tests, the master execution script reads in the configuration files (specified in the YAML format), identifies the corresponding reader or test classes to load, and passes the configurations to the class and executes the class methods.
The master execution script has access to all available catalogs and tests, and can execute all desired combinations.
By default, the master execution script is set to run periodically, thereby automating the full validation process.

\rev{The master execution script is also equipped to handle failures, relying on Python's context manager. The execution script captures all exceptions and traceback information, together with any content printed to standard output or standard error during runtime, and stores them in a log file. This is done for each of the combinations of all tests and catalogs, and when one of them fails, the master execution script writes out the log file, and continues to the next combination without being interrupted. }

This design makes our framework easily expandable. Including new catalogs or validation tests requires no changes to existing code. Once the new reader or test class and its corresponding configuration file are placed in the pre-specified location, the master script will automatically include the new catalog or test in future runs. 
In this fashion, catalog readers and validation tests can remain agnostic about what catalogs and tests are available, and hence do not require updates when new catalogs and tests are added.

\rev{For the seven tests and eight catalogs presented in \autoref{sec:validation}, since the catalogs are fairly small (made out of the 100\,Mpc$\,h^{-1}$ box), it takes only about 10 minutes on a single CPU core to run all of the combinations. The eight catalogs take less than 25\,GB of disk space (excluding the underlying DMO simulation). If we were to run a test whose computational cost scales linearly with the number of galaxies on a mock catalog that corresponds to about 10,000 square degrees, it would take tens of minutes on a single CPU core, and this catalog would take about 1TB of disk space (depending on how many galaxy properties are stored). These numbers are still manageable by modern standards; however, in the future when facing very large catalogs and more computationally involved tests (such as two-point statistics), we will need distributed computation. Although we have not yet explored this direction, we believe our framework is flexible enough to accommodate distributed computation. In particular, our framework can be configured to run different tests on different catalogs using separate cores, without the need for cross-node communication (i.e., embarrassingly parallelizable).}

All of the results (including the plots and summary statistics generated by the validation tests) can be archived periodically.
Although the user can certainly inspect individual output files to access validation results, this process will rapidly become tedious with the increasing numbers of catalogs and tests. To avoid this difficulty, we have built a Web interface\footnote{\https{portal.nersc.gov/project/lsst/descqa}} at NERSC to assist users in quickly inspecting validation results.
When users visit the Web interface, the Common Gateway Interface scripts will read in the output files available on the filesystem, and present users with a visual summary of the results. 
The Web interface also allows the user to browse through different runs. When each run is executed, a copy of the code used is recorded so that the results can be easily tracked. 

\begin{figure*}[tb!]
\centering \includegraphics[width=\textwidth,trim=0.8in 5.2in 0.4in 0.8in,clip]{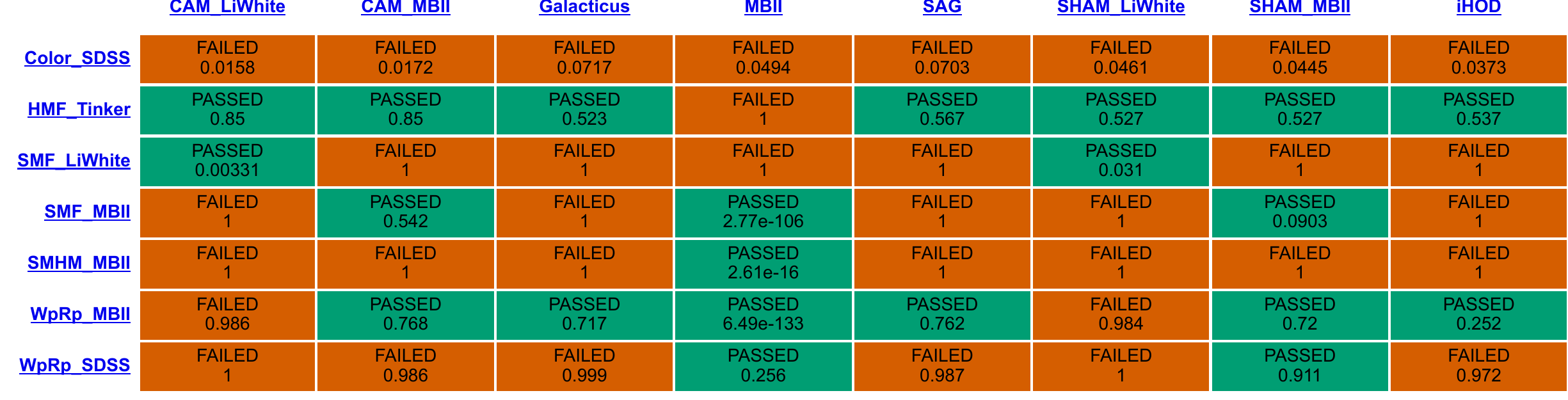}
\caption{\label{fig:web}%
Screenshot of the summary page of the DESCQA Web interface$^\text{\ref{fn:this-run}}$ demonstrating one aspect of the framework. For the current set of catalogs and validation tests, this matrix provides a quick overall view of how each catalog (column) performs on each validation test (row).
Each validation test also provides a score for each catalog it runs on and a passing threshold that is designed to indicate whether the catalog being tested meets the specific requirement of the test under consideration.
For the current set of validation tests, the scores are always between 0 and 1, and a larger score value indicates a larger discrepancy between the catalog in consideration and the validation data set{; note that the specific values and passing criteria are for demonstration purposes and do not reflect the actual LSST~DESC requirements.}
This matrix enables users to quickly identify the synthetic catalogs that satisfy their needs. A detailed discussion of these validation tests is provided in \autoref{sec:validation}. 
}
\end{figure*}

\autoref{fig:web} shows {an example of} the summary page of the Web interface, which is presented in the form of a validation matrix. This matrix provides a quick summary of all validation tests (rows) and all available catalogs (columns). Each colored cell shows the corresponding test result. 
{Each of the current set of validation tests provides a score that is between 0 and 1. A higher score indicates a larger discrepancy between the catalog under consideration and the validation data set. When the score is higher than a certain predefined value, it is noted as ``failed.'' As already noted, the specific values and passing criteria are for demonstration purposes and do not reflect the actual LSST~DESC requirements.
In this matrix view,} users can further click on the header or the cell to see the associated plots and output files. 
{This interface helps the users} to quickly find the catalogs that satisfy their desired requirements.

The DESCQA framework utilizes the filesystem to serve the Web interface and avoids direct communication between the test scripts and the Web interface. Hence, the framework can be easily adapted and applied more broadly to many other comparison tasks, such as a comparison of different code implementations.

\rev{
\subsection{Documentation and Maintenance}

The DESCQA framework faces many different types of users, and hence requires many different forms of documentation. The framework is highly modularized, enabling different types of users to contribute without the overhead of understanding the full framework. 
Here we describe the different requirements on and our implementation of the documentation as well as related issues regarding maintenance.

\begin{enumerate}
\item For Web interface users who want to browse the results and plots of validation tests: the Web interface is self-explanatory and requires little documentation. On the front page, we provide basic instructions of how to navigate the interface, along with links to our papers, code repositories, and internal documentation pages, where users can easily find further information.

\item For catalog users who want to access the synthetic catalogs through the reader interface: we provide both API documentation\footnote{\label{fn:gcr}\https{yymao.github.io/generic-catalog-reader/}} to the reader interface and also an example code\footnote{\https{github.com/LSSTDESC/gcr-catalogs/tree/master/examples}} (many of which use easily browsable Jupyter notebooks) to help users understand quickly how to access the synthetic catalogs using the reader interface. 

\item For users who want to implement a validation test to be included in the DESCQA framework: as discussed above, since the validation tests themselves are very much independent of the rest of the framework, the knowledge required for a test writer to be able to contribute is much reduced. In particular, test developers need only to implement a subclass that inherits the base class of validation tests and to follow the instructions\footnote{\https{github.com/LSSTDESC/descqa/tree/master/descqa}} on how to implement a few specific member methods. We also provide step-by-step instructions on how to manually trigger the validation framework to test newly implemented validation tests\footnote{\https{github.com/LSSTDESC/descqa/blob/master/README.md} and \https{github.com/LSSTDESC/descqa/blob/master/CONTRIBUTING.md}}. 

\item For catalog providers who want to contribute their catalogs: similar to the case for new test developers, new catalog providers need only to implement a subclass that inherits the reader base class and to follow the instructions$^\text{\ref{fn:gcr}}$ on how to implement a few specific member methods. The catalog providers can test their newly implemented readers either with the full DESCQA framework or through the importable reader module. 

\item For users who maintain the Web interface and execution scripts, which includes DESCQA framework maintainers but not most regular users: we have made these components of our framework highly modularized and self-documenting such that future maintainers can navigate the code easily. We are working on other visual aids such as flowcharts to help future maintainers understand the code structure better. 
\end{enumerate}

While we continue to improve these various aspects of documentation, recent feedback from users both within the DESC Collaboration and elsewhere suggest that we already have adequate documentation for the different types of users to utilize or to contribute to this framework. 

In addition, we work closely with the computing infrastructure working group of the Collaboration to ensure that the code base of this framework is kept up to date with the development environment (e.g., Python and Python packages), so as to reduce potential dependencies on deprecated packages and to benefit from better performance and new features. 
The Collaboration intends to continue to use, support, and develop the DESCQA framework, and will help to ensure that a period of overlap and effective communication is enabled between the current and future maintainers.

}


\section{Case Studies}
\label{sec:validation}

{To demonstrate the design and features of the DESCQA framework, we present five validation tests as case studies. \autoref{tab:tests} provides a summary of these five tests and the corresponding validation data sets and criteria. 
These criteria can be defined to satisfy specific science goals; however, as LSST~DESC is still finalizing science requirements, the criteria used here are for demonstration purposes.}

{As mentioned earlier, one important requirement of this framework is that it needs to be suitable for a wide range of heterogeneous synthetic catalogs. 
Hence, we select eight realistic synthetic catalogs that encompass the major classes of methods that are generally used to create synthetic galaxies (HOD, SHAM, SAM, and hydrodynamical simulations) to use in our case studies.}

{\autoref{tab:models} summarizes the eight catalogs used here. In particular, the hydrodynamical galaxy catalog is extracted from the MassiveBlack-II (MBII) simulation, further described in \autoref{sec:mbII}. All other catalogs are built upon the dark structures of the same gravity-only simulation, MBII DMO, which is a companion run of MBII using the same initial conditions, resolution, and box length as its hydrodynamical counterpart. The description of MBII DMO can also be found in  \autoref{sec:mbII}. The details of how each catalog is implemented can be found in the rest of the subsections of \autoref{sec:catalogs}.}

In order to keep the creation of the catalogs as simple as possible, we choose to compare them only at a single snapshot (fixed redshift). This approach is sufficient to establish the functionality of the DESCQA framework. In the near future, we will expand the framework to include light-cone catalogs.

{For each test, we present a summary comparison plot to compare the results from different catalogs and the validation data set, and discuss how these case studies have in turn influenced the design of the framework. We summarize these findings in \autoref{sec:findings}.}
All plots presented here are directly taken from the output of the DESCQA framework, without any further editing. 
We encourage the reader to further investigate all results directly using the DESCQA Web interface,\footnote{\label{fn:this-run}\weburl{} is where this particular run locates} which includes some tests that we did not display here and also includes plots that compare the results for each catalog with validation data separately.

\begin{table*}[tb!]
\begin{center}
\caption{Summary of Validation Tests and the Corresponding Interim Validation Data Sets and Passing Criteria Presented in this Study.\label{tab:tests}}
\begin{tabular}{l|lll}
Test & Section & Validation Data Sets & Passing Criteria \\
\hline
Stellar mass function (SMF) & \ref{sec:smf} & \cite{LiWhi09}, MBII (not shown) & $\text{CDF}(\chi^2, \text{dof}) < 0.95$ \\
Halo mass function  (HMF) & \ref{sec:hmf} & \cite{Tinker2008} & $\text{CDF}(\chi^2, \text{dof}) < 0.95$ \\
Stellar mass--halo mass (SMHM) relation  & \ref{sec:smhm} & MBII & $\text{CDF}(\chi^2, \text{dof}) < 0.95$\\
Projected two-point correlation function & \ref{sec:wprp} & SDSS \citep[as in][]{Reddick2013}, MBII (not shown) & $\text{CDF}(\chi^2, \text{dof}) < 0.95$\\
Galaxy color distribution & \ref{sec:color} & SDSS & $\omega < 0.05$ for all 4 colors (shifted)\\
\end{tabular}
\end{center}
\end{table*}

\begin{table*}[tb!]
\begin{center}
\caption{Summary of the Synthetic Catalogs Used in the Case Studies. See \autoref{sec:catalogs} for Details. \label{tab:models}}
\begin{tabular}{l|llll}
Catalog & Abbreviation & Appendix & Model Type & Variants \\
\hline
MassiveBlack-II Hydrodynamic & MBII & \ref{sec:mbII-galaxy} & hydrodynamic simulation & --\\
Improved Halo Occupation Distribution & \ihod{} & \ref{sec:ihod} & halo occupation distribution & --\\
SHAM-ADDSEDS & SHAM & \ref{sec:AMM} & abundance matching & LiWhite, MBII\\
Conditional Abundance Matching & CAM & \ref{sec:CAMM} & abundance matching & LiWhite, MBII\\
Semi-Analytic Galaxies & SAG & \ref{sec:SAG} & semi-analytic model & -- \\
Galacticus  & Galacticus & \ref{sec:gal} & semi-analytic model & -- \\
\end{tabular}
\end{center}
\end{table*}

\subsection{Stellar Mass Function}
\label{sec:smf}
\begin{figure}[tb!]
\centering\includegraphics[width=\columnwidth]{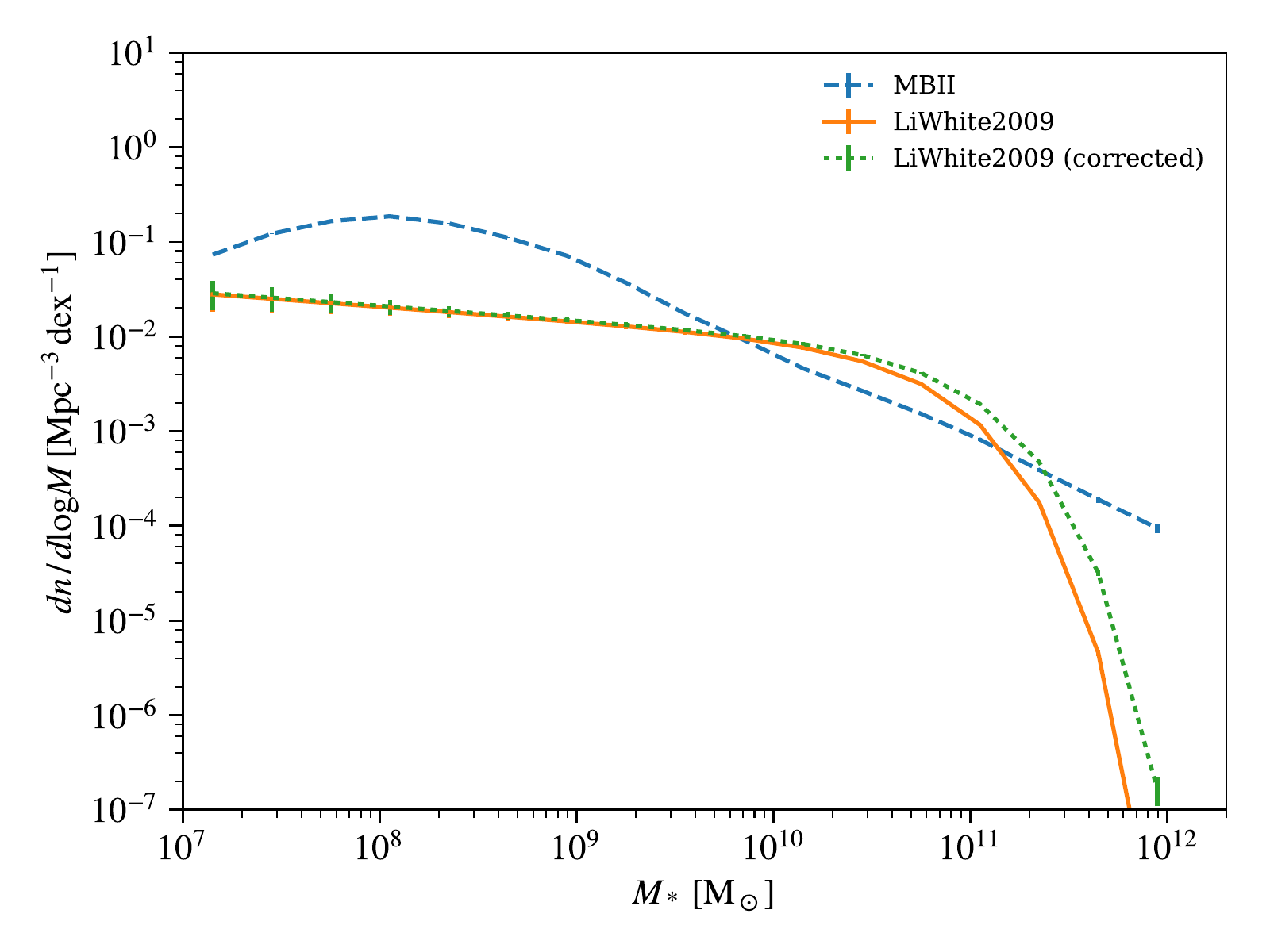}
\caption{\label{massf}
Stellar mass function from the MBII simulation (blue dashed line) and from \citet[orange solid line]{LiWhi09}, both at $z=0.06$. The green dotted line shows the \cite{LiWhi09} stellar mass function but corrected for the mass in stellar remnants, as described in \autoref{sec:smf}, to ensure that its stellar mass definition is consistent with that of MBII.
To build the abundance-matching-based catalogs, we use both
the MBII stellar mass function and the stellar-remnant--corrected measurements from \cite{LiWhi09} as input.}
\end{figure}

For each of the synthetic catalogs, we calculate the stellar mass density as a function of the total stellar mass for each galaxy. The densities are derived from the number counts of galaxies in each stellar mass bin, divided by the simulation volume. These densities are compared with the stellar mass functions from the MBII hydrodynamic simulation and from \cite{LiWhi09}. \autoref{massf} shows a comparison between these two stellar mass functions. 

Stellar masses are most commonly defined as the mass locked up in long-lived stars and stellar remnants. {However, synthetic catalogs and the validation data sets may all have defined stellar mass differently, and the discrepancy cannot be homogenized by the reader interface}. For the SAM models, the total stellar mass is the sum of the disk and spheroid components. {For the SHAM-based models, the stellar masses correspond to the galaxy catalogs to which they were matched.
On the other hand,} the stellar masses used to construct the \cite{LiWhi09} stellar mass function are taken from the New York University Value-Added Galaxy Catalog~(NYU-VAGC; \citealt{2005AJ....129.2562B}), and were derived from the \textsc{kcorrect} code \citep{blanton2007}. As such, they do not include the mass in stellar remnants (white dwarfs, neutron stars, etc.), which is more commonly included in the definition of stellar mass. For a \cite{chabrier_galactic_2001} initial mass function and using stellar data from \cite{marigo_chemical_2001} and \cite{portinari_galactic_1998}, we find that the fraction of the original population mass in stellar remnants for a single-age population of age 10\,Gyr is 14.6\%, with 39.7\% of the original population's mass remaining in stars after this time. Therefore, we shift the \cite{LiWhi09} mass function masses by $+0.136$ (i.e., $\log\,[39.7\%/(39.7\%+14.6\%)]$) in order to include the mass in stellar remnants. 

Estimates of the stellar masses of galaxies also suffer from other sources of systematic error. For example, \cite{mobasher_critical_2015} show that uncertainties arise from the template-fitting procedures used to estimate stellar masses from multiband photometry. Although they considered other surveys, they demonstrated that systematics at the 0.1\,dex level can arise from these error sources. In the specific case of the stellar mass function from the Sloan Digital Sky Survey (SDSS), \cite{Bernardi2013} reanalyzed the SDSS photometry and concluded that there are significant systematic biases in the inferred stellar masses arising from the choice of surface brightness profile fit to the data. In their latest work, \cite{bernardi_high_2016} estimate that the photometric systematic errors are at the level of 0.1\,dex. Here we have chosen the \cite{LiWhi09} measurement as an example, which is significantly different from the measurement in MBII.

As mentioned in \autoref{sec:descqa}, each test must provide a summary score. For this test, the summary score is the probability for a $\chi^2$ distribution, given the number of bins, to have a value less than the one calculated from the comparison of the catalog result to the validation data:

\begin{equation}
\chi^2 = \sum_{i,j} \left(\phi_i - \hat{\phi}_i\right) \left[\left(C + \hat{C}\right)^{-1}\right]_{i,j} \left(\phi_j - \hat{\phi}_j\right),
\end{equation}
where $\phi_i$ and $\hat{\phi}_i$ are the differential stellar mass number density for mass bin $i$ calculated from the catalogs and from the validation data, respectively, $C$ is the covariance matrix calculated from the catalog using the jackknife resampling method, and $\hat{C}$ is the covariance matrix calculated from the validation data, in which case we include only the diagonal terms. 
To evaluate $C$, we implement the jackknife resampling method by dividing the simulated box into $5^3=125$ smaller cubic boxes.

\begin{figure}[tb!]
\centering\includegraphics[width=\columnwidth]{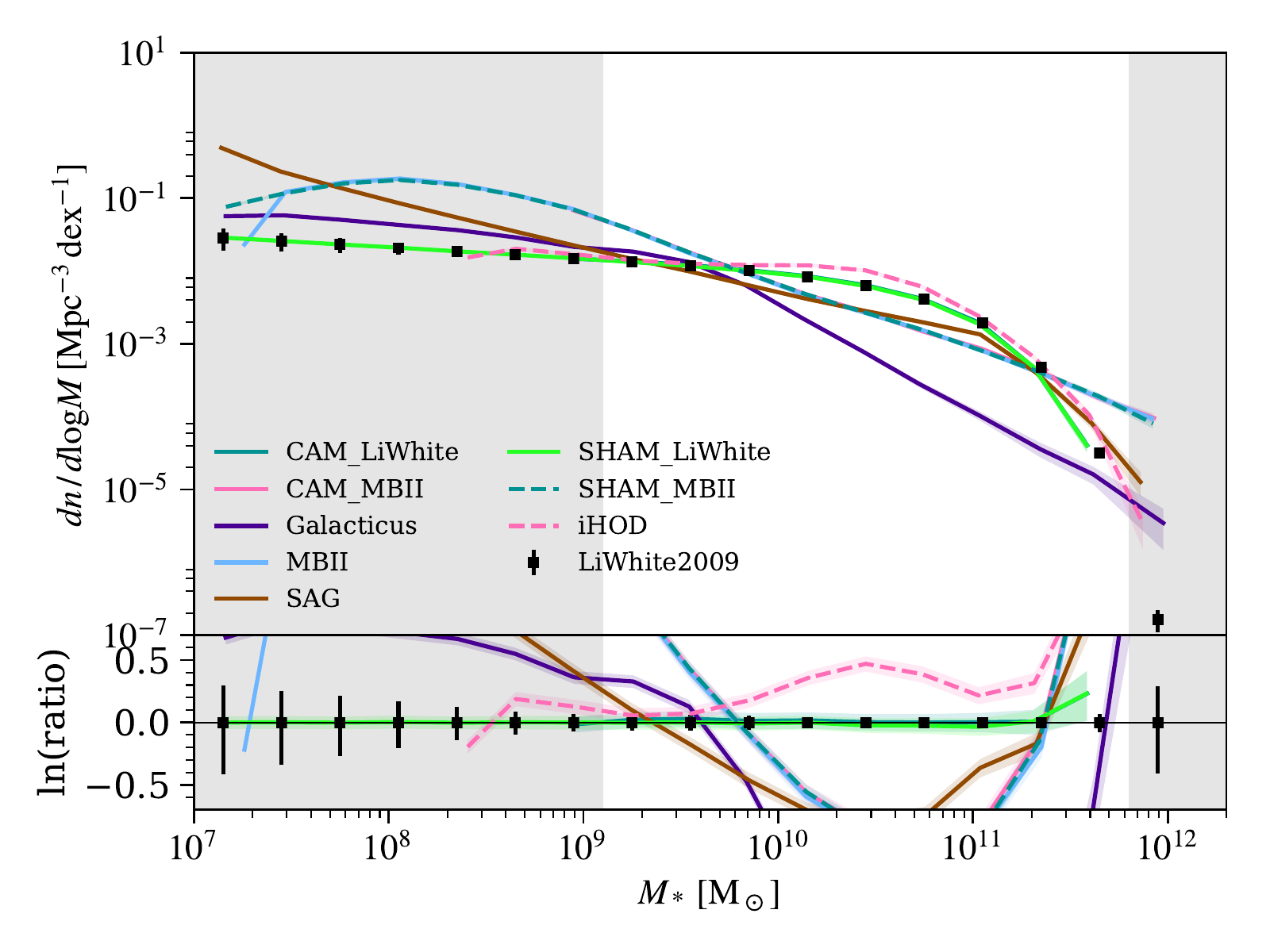}
\caption{\label{fig:smf_LiWhite_comp}%
Stellar mass functions for all synthetic galaxy catalogs (shown as the colored lines with shaded regions) compared to the validation data set (shown as black points with error bars). In this case, the validation data set is the stellar-remnant-corrected \citet{LiWhi09} stellar mass function.
The labels in the legend indicate the method used to generate the catalog (e.g., SHAM\_LiWhite is the SHAM-ADDSEDS method described in \autoref{sec:AMM} tuned to the \citet{LiWhi09} stellar mass function).
The shaded colored region shows the square root of the diagonal terms of the covariance matrix $C$ of the stellar mass function, obtained from the jackknife resampling method, for the synthetic catalog of the corresponding color.
For a close-in comparison, the lower panel shows the natural logarithm of the ratio to the validation data set (i.e., $\ln\,y/y_\text{ref}$). 
To see the result for each catalog more clearly, please visit the Web interface$^\text{\ref{fn:this-run}}$. 
Points within the vertical gray bands are not used to calculate the scores that appear in \autoref{fig:web}.}
\end{figure}

The criterion to pass this test is set to be a score less than 0.95 (equivalent to having a right-tail $p$-value larger than 0.05). 
{When designing this test, we also notice that in most cases, the passing criterion may not apply to the full range of stellar masses. For example, the low-mass end is bound to be affected by resolution, and, depending on the user's application, this may or may not be an issue. 
Hence, we design the test to have a configurable validation range. We demonstrate this feature here (as the white band in \autoref{fig:smf_LiWhite_comp}) and require the synthetic catalogs to reproduce stellar masses above $10^9\,\msun$. We also exclude the most massive bin when calculating the score as the last bin is dominated by cosmic variance.}

\autoref{fig:smf_LiWhite_comp} shows the results for the stellar mass function compared to the observational measurements from~\cite{LiWhi09}. The reader is encouraged to inspect the results in more detail with the help of the DESCQA Web interface.  The validation data is shown in black solid circles with error bars, and the synthetic catalogs are represented by the colored lines with shaded error bands.

By construction, CAM\_LiWhite and SHAM\_LiWhite are almost identical to the \cite{LiWhi09} validation data set as they are based on the abundance-matching technique, which guarantees an exact match to the input stellar mass function. 
For the same reason, CAM\_MBII and SHAM\_MBII are very close to the MBII stellar mass function.
\ihod{} was originally tuned to fit \cite{LiWhi09} as well, leading to good agreement in this test.  More interesting are the results from the two SAM approaches. Both of them overpredict the stellar mass function at low masses compared to the \cite{LiWhi09} measurement, SAG somewhat more than Galacticus, similar to MBII. The shape of the stellar mass function for the SAMs does show a hint of a knee, similar to what is seen in the validation data set and unlike the MBII catalog, but the shape of the measurement is still not captured very well. 
{This test demonstrates that, if a catalog user wants to impose a stringent requirement such as the one we use here, currently only SHAM-based models would pass the test due to their construction method. Hence, careful consideration is advised when designing the requirement for SMF tests.}

\subsection{Halo Mass Function (HMF)}
\label{sec:hmf}
The mass distribution of halos is one of the essential components of precision cosmology and occupies a central place in the paradigm of structure formation.
There are two common ways to define halos in a simulation.
One of these, the spherical overdensity definition, is based on identifying overdense regions above a certain threshold.
The threshold can be set with respect to the (time-varying) critical density $\rho_c = 3 H^2 / 8 \pi G$ or the background density $\rho_b = \Omega_m \rho_c$.
The mass $M$ of a halo identified this way is defined as the mass enclosed in a sphere of radius $r_\Delta$ whose mean density is $\Delta \rho_c$, with common values ranging from 100 to 500.
The other method, the friends-of-friends (FOF) algorithm, is based on finding neighbors of particles and neighbors of neighbors as defined by a given separation distance \citep[][]{Einasto1984, Davis1985}.
The FOF algorithm is essentially an isodensity estimation method (mass enclosed within a given isodensity contour). FOF halos can have arbitrary shapes, since no prior symmetry assumptions have been made; the halo mass is simply the sum of the masses of particles that are halo members.

Here, we calculate the HMF from each catalog for the distinct halos and provide a comparison to the well-established analytic fit by \citet{Tinker2008} for spherical overdensity-defined halos ($M_\text{100c}$), which is accurate at the 5--10\% level at $z=0$ for a $\Lambda$CDM cosmology.
We have also implemented (not shown) the \citet[]{1999MNRAS.308..119S} and \citet{Bhattacharya2011} fits for the FOF halos, in addition to many other analytic mass function fits; for details, see~\citet[]{Lukic2007}.
The original code was written in Fortran, and we provide a simple Python interface to include the code in DESCQA; we have made the code publicly available\footnote{\https{github.com/zarija/HaloMassFunction}}.
This test uses the same summary statistic as the SMF test as described in \autoref{sec:smf}, except that the covariance of the validation data $\hat{C}$ is set to Poisson errors for the diagonal terms and zeros for the off-diagonal terms for this particular test, as the validation data is an analytic fit.

\begin{figure}[tb!]
\centering\includegraphics[width=\columnwidth]{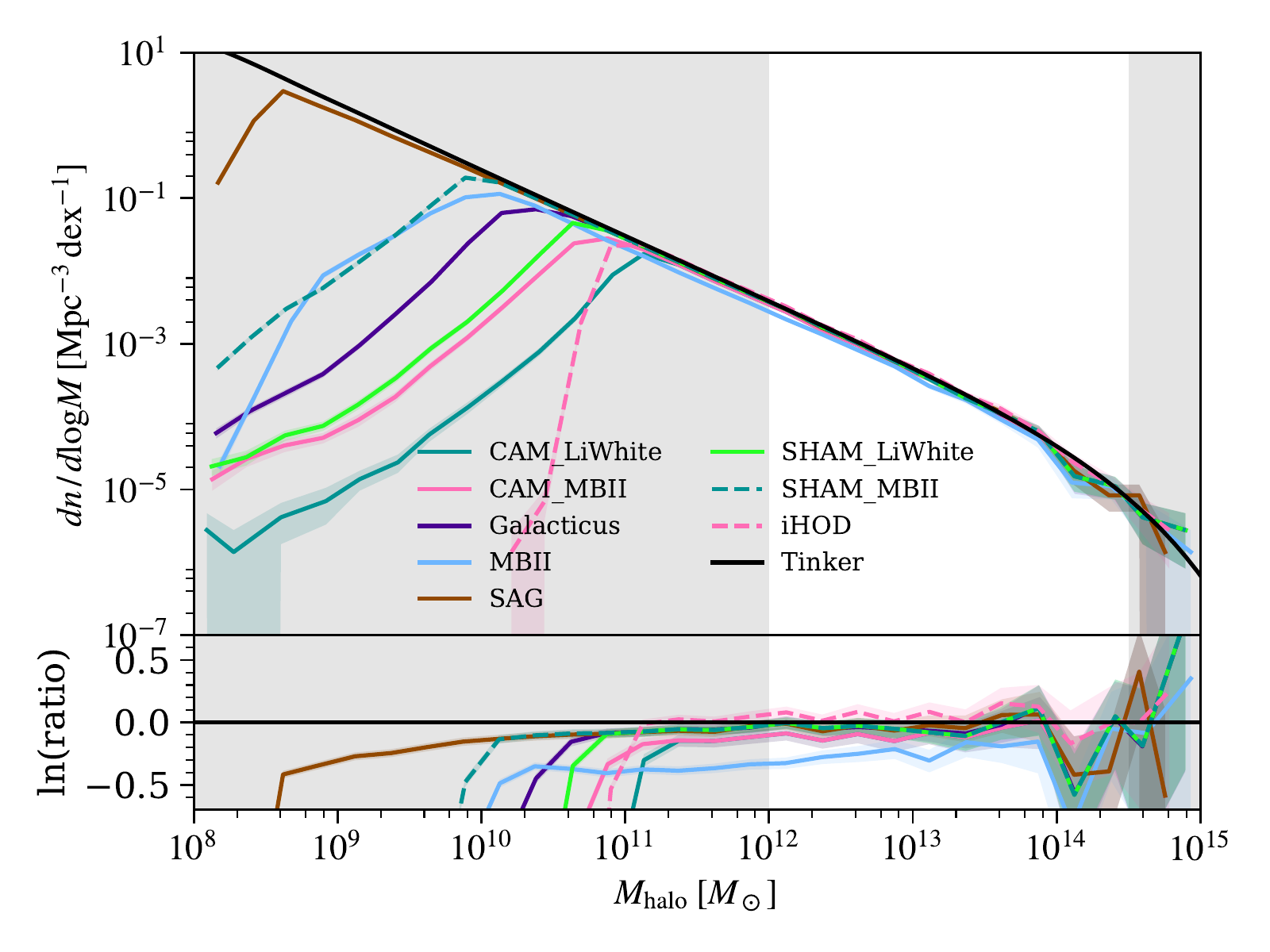}
\caption{\label{fig:hmf}%
Same as \autoref{fig:smf_LiWhite_comp} but for the halo mass function. Here, the validation data set is an analytic fit from \cite{Tinker2008} and is shown as the black line.} 
\end{figure}

{To compare the mass function for only distinct (host) halos from each catalog, this test requires two important pieces of information: (1) the halo mass associated with each galaxy and (2) whether or not the galaxy is a central galaxy. We then select only halos that are associated with central galaxies. Hence, although all the interim catalogs presented here use the same base simulations, catalogs that assign central galaxies differently may have different HMFs. 
We note that the HMF we evaluate here is defined slightly differently from the usual HMF in the sense that we require the halos to host central galaxies. Distinct halos that do not contain any central galaxy are not included in these catalogs. As a result, different catalogs include very different abundances of low-mass halos, depending on their halo occupation function at the low-mass end. 
This effect can be clearly seen in~\autoref{fig:hmf}: above $\sim10^{11}\,\msun$, all mass functions agree extremely well and follow the Tinker fit at the expected level of accuracy. Below this mass, the catalogs start to disagree.
These issues should be taken into account when designing the requirement for the HMF test.
}

{Other possible discrepancies that cannot be homogenized by the reader interface include cosmology, halo mass definitions, and halo finders.
Different research groups often use different halo mass definitions and halo finders, and hence flexibility in this regard is important. Our test routine provides different fitting functions that can be chosen to match the underlying cosmology and halo mass definition used to create the synthetic catalog. 
Differences in halo finders, on the other hand, are more difficult to deal with. For example, the MBII hydrodynamic simulation-based catalog was generated using an FOF halo finder with linking length of $b=0.2$, while all other synthetic catalogs that we used in this study are based on the same $N$-body simulation (MBII DMO) analyzed with the \textsc{Rockstar} halo finder and the same spherical overdensity definition.
In \autoref{fig:hmf} we see that the MBII HMF is overall lower relative to results from the populated catalogs, in particular for the low-mass halos.
This result is not only due to different halo mass definitions, but also due to the presence of baryons as indicated by the findings by~\cite{2015MNRAS.453..469T}, where the MBII HMF was compared to the DMO mass function (for the same linking length $b=0.2$) and a difference at the 15\% level was found for halos with masses of $\sim 10^{11}\,\msun$. Similar results were found in the OverWhelmingly Large Simulations \citep[OWLS;][]{2014MNRAS.442.2641V}, and in the Magneticum simulations \citep{2016MNRAS.456.2361B}. This shows that baryonic effects and halo definitions are potentially important for designing the requirements for an HMF test. 
}

\subsection{Stellar Mass--Halo Mass (SMHM) Relation}
\label{sec:smhm}
The SMHM relation, defined as the stellar mass of central galaxies within a distinct halo of total spherical overdensity mass, is a now well-established estimator of the efficiency of gas cooling and star formation over a wide range of halo masses. Both observational and theoretical works have shown that the efficiency of the stellar mass assembly (M$_*$/M$_\text{halo}$) peaks at the scale roughly corresponding to the knee of the stellar mass function and declines at smaller and larger masses. This is generally interpreted in simple terms as the efficiency of the stellar mass assembly being suppressed at the low-mass end by supernova feedback and at the high-mass end by active galactic nucleus (AGN) feedback \citep{2012RAA....12..917S}. 

Similar to the HMF test, the SMHM relation is not directly comparable to observational data. Not all validation tests, however, need to be tests that compare synthetic catalogs with actual observations, as some tests are designed to aid catalog users in understanding the features {and characteristics} of each catalog, or to provide comparisons with other results in the literature, or from other models. {For example, when a user validates the SMHM relation, they might actually be verifying if the catalogs match a specific SMHM relation that is derived from a hydrodynamical simulation or inferred from a theoretical model. Moreover, the user can also validate only a certain regime of the SMHM relation, for example, to verify if the effect of AGN feedback is present in the catalog under consideration.}

In the current DESCQA test implementation, we use the results from the MBII hydrodynamical simulation as {an interim} validation data set. As mentioned in \autoref{sec:mbII} for MBII, the halos were identified with an FOF halo finder  with linking length $b=0.2$. The halo definition is therefore different for the validation data set than for the synthetic catalogs, where the mass definition is the virial mass ($M_\text{vir}$). In addition, due to baryonic effects, the halo masses from the MBII hydrodynamic simulation are lower than those from the DMO simulation (see \citealt{2015MNRAS.453..469T} and \autoref{sec:hmf}). 
The SMHM relation uses only distinct halos (host halos) and excludes subhalos.
{As such, most of the caveats that apply in the HMF test would also apply here.}
This test also uses the same summary statistic as the stellar mass function test as described in \autoref{sec:smf}, and the same validation range as the HMF test (\autoref{sec:hmf}). 

\begin{figure}[tb!]
\centering\includegraphics[width=\columnwidth]{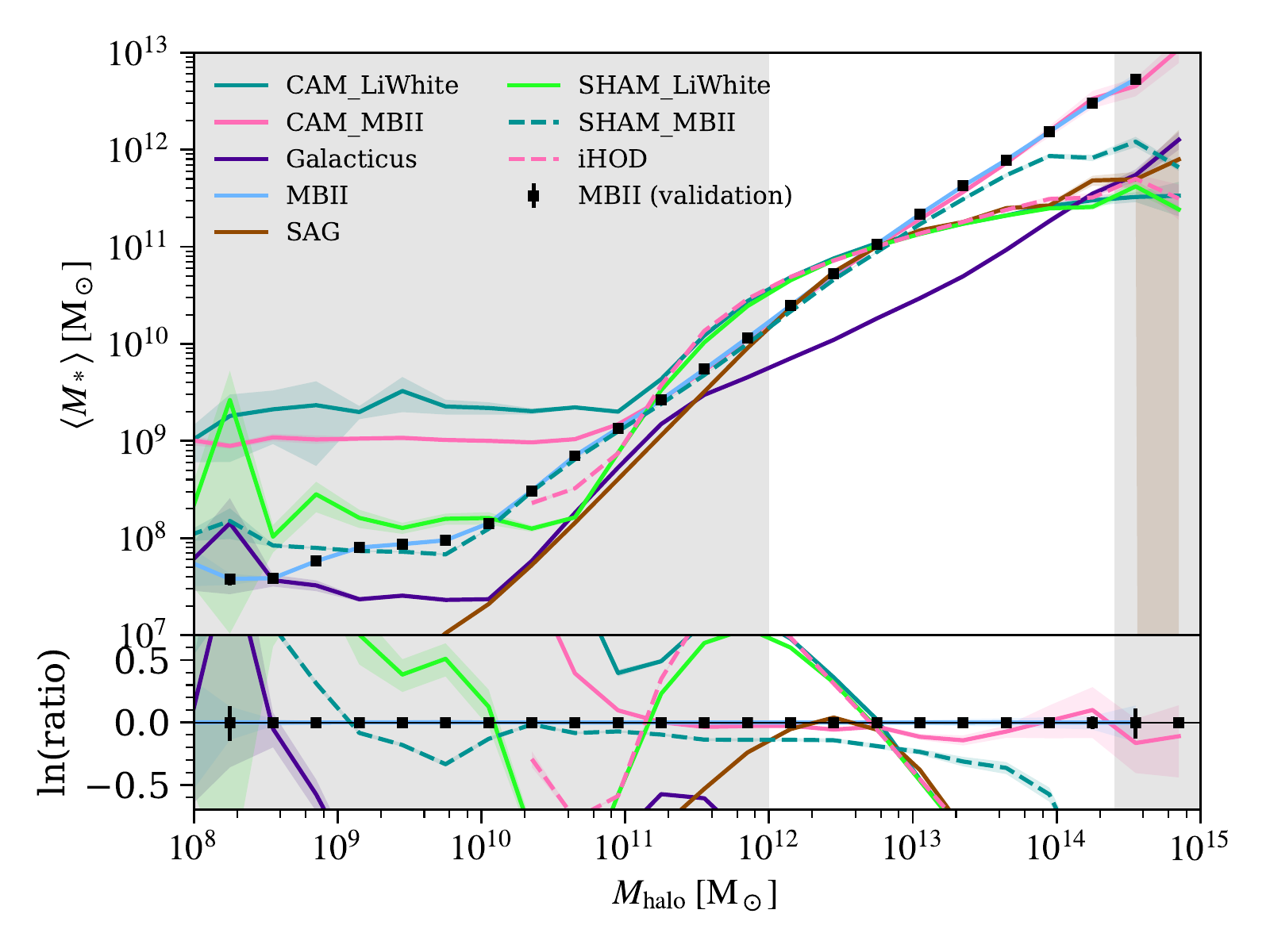}
\caption{\label{fig:smhm_mb2_comp}%
Same as \autoref{fig:smf_LiWhite_comp} but for the stellar mass--halo mass relation. Here, the validation data set (black points) is drawn from the MBII hydrodynamical simulation, and hence the MBII catalog shows a perfect match by construction.}
\end{figure}

\autoref{fig:smhm_mb2_comp} shows the results from MBII and the catalog results with error bands. The MBII result is trivially perfect by construction because it is a self-comparison. SHAM\_MBII and CAM\_MBII perform quite well over most of the mass range as these models were tuned to the MBII stellar mass function. However, at the high-mass end, the SMHM relation from SHAM\_MBII flattens out due to the constant scatter used in the abundance-matching technique, while MBII's SMHM relation exhibits a much smaller scatter in stellar mass at high halo mass.
SHAM\_LiWhite and CAM\_LiWhite perform reasonably well over the intermediate-mass range where the MBII and Li \& White stellar mass function are closest. The overprediction of MBII at the low- and high-ass end compared to Li \& White is reflected in the discrepancy seen in SHAM\_LiWhite and CAM\_LiWhite. Galacticus is overall lower than MBII but has the correct rise at the high-mass end. The SAG catalog underpredicts the SMHM relation compared to the MBII test for the low-mass halos. \ihod{} overall fits reasonably well though the results are worse at extreme mass values. 
{Note that, except for MBII (which is a self-comparison), none of the catalogs passes the current validation criterion. This indicates that more thoughtful criteria and more realistic validation data sets (e.g., ones derived from empirical models) should be adopted if catalog users view the SMHM relation as an essential measurement that catalogs must reproduce.}

\subsection{Projected Two-point Correlation Function}
\label{sec:wprp}
{The projected galaxy two-point auto-correlation function, $w_p(r_p)$, is one of the most-used clustering statistics for testing both cosmology and galaxy formation models. Here we describe our test to compare $w_p(r_p)$ among different synthetic catalogs and against observational and simulated data. Since our interim synthetic catalogs are given at a single epoch,} we calculate $w_p(r_p)$ using the thin-plane approximation. 
We use the catalogs at one epoch and then add redshift space distortions along one spatial axis ($z$-axis).
We then calculate the projected pair counts, with a projection depth of $\pm\, 40\,h^{-1}$Mpc.  We assume periodic boundary conditions for all three spatial axes. 

This test also uses the same summary statistic as the stellar mass function test described in \autoref{sec:smf}, though the evaluation of the covariance for the catalog, $C$, is slightly different, and we include the full covariance of the validation data, $\hat{C}$. 
We estimate the sample variance of $w_p(r_p)$ using the jackknife technique. We divide the projected 2D plane ($x-y$ plane) into $10 \times 10$ smaller regions, with each region having an area of $(10 \, h^{-1}\text{Mpc})^2$. We then re-evaluate the $w_p(r_p)$ when removing one region at a time.
The code that calculates $w_p(r_p)$ and its jackknife variance is publicly available\footnote{Module `CorrelationFunction' in \https{bitbucket.org/yymao/helpers}}.

In this paper, we compare the $w_p(r_p)$ from each catalog for all galaxies that have a stellar mass larger than $10^{9.8} \, h^{-2}\msun$.
We use two interim validation data sets for this test: the $w_p(r_p)$ calculated from the MBII hydrodynamical simulation and the $w_p(r_p)$ from SDSS as presented in \citet{Reddick2013}. This measurement was made on the volume-limited samples
from the NYU-VAGC catalog~\citep{2005AJ....129.2562B}, based on Data Release 7 from the SDSS \citep{Padmanabhan2008,Abazajian2009}.

\begin{figure}[tb!]
\centering\includegraphics[width=\columnwidth]{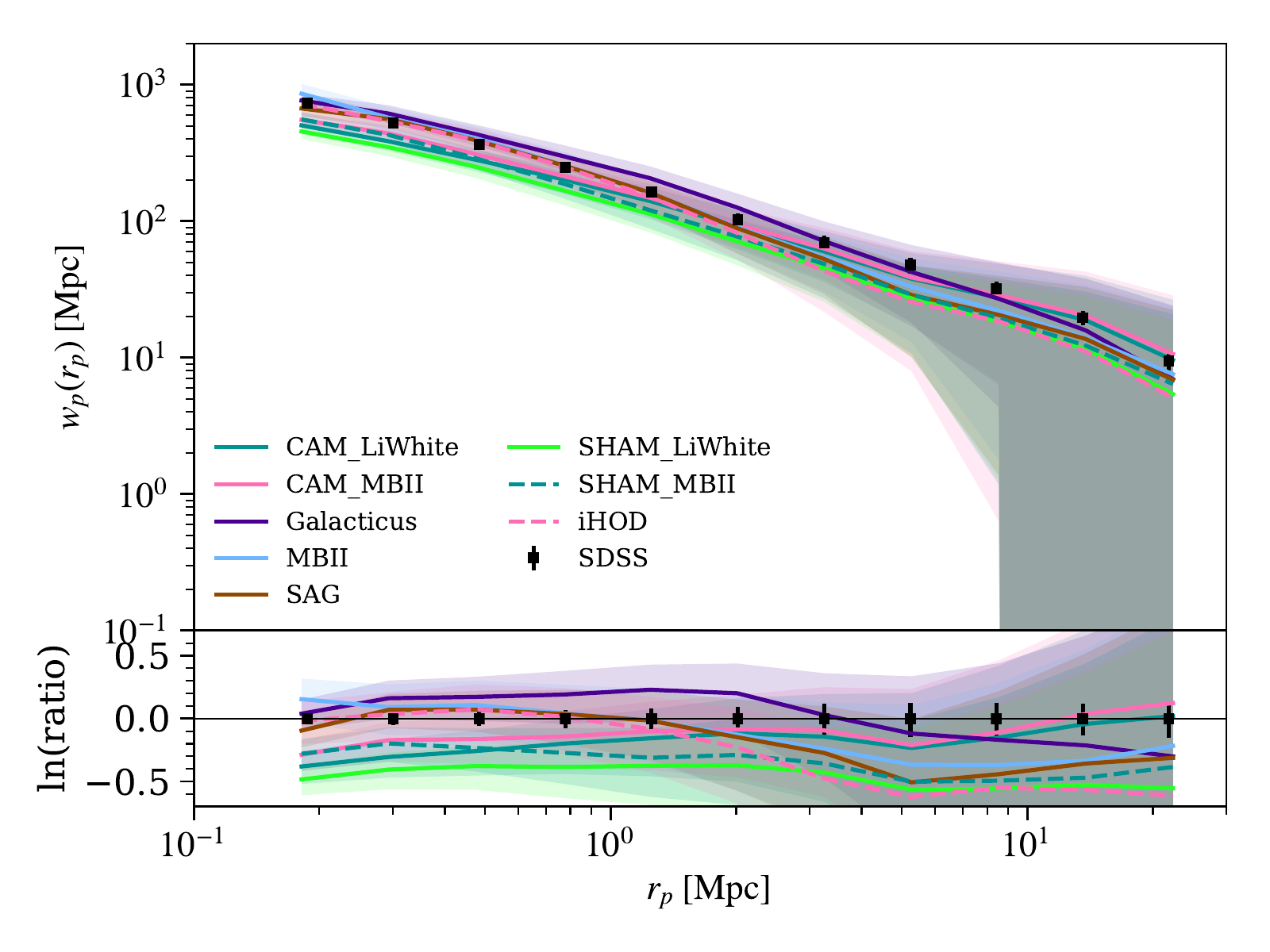}
\caption{\label{fig:wprp}%
Same as \autoref{fig:smf_LiWhite_comp} but for the projected two-point correlation function for all galaxies down to a stellar mass of $1.28 \times 10^{10} \,\msun \; (=10^{9.8}\,h^{-2}\msun)$ . Here the validation data set (black points) is from SDSS as presented in \citet{Reddick2013}. Visit the Web interface to see the individual error bars more clearly.}
\end{figure}

{In \autoref{fig:wprp}} we show the results of comparisons with the SDSS data being used for validation; the reader is encouraged to inspect the comparisons with MBII as the validation data directly on our Web interface.
Overall, the agreement between the catalogs and SDSS data is rather good{, and most catalogs pass our preliminary validation metric}. Most synthetic catalogs underpredict the small-scale clustering when compared with SDSS data, though they are still within the 2$\sigma$ errors. Due to the small volume of our simulation box, the jackknife sample variance dominates the error budget. The data are likely to better distinguish between the same models if they are applied to a larger cosmological volume.
{Hence, one should consider including the volume of the catalogs as part of the evaluation for a $w_p(r_p)$ test.}

\subsection{Galaxy Color Distribution}
\label{sec:color}
We also include a test of how well the synthetic galaxy color distributions compare to the observed colors of galaxies. In principle, this test should be done with light-cone catalogs that cover the same redshift range as the observed data set, with the same set of cuts on observed properties. However, {for proof of concept, we present it using our current catalogs at a single epoch}, $z=0.0625$ (except for the SAG catalog, which is at $z=0$, and the Galacticus catalog, which is at $z=0.05$).

We determine the color distributions in these catalogs {for comparison with our validation data set ---} measurements of the $ugriz$ colors of a volume-limited sample of $0.06<z<0.09$ galaxies from SDSS DR13 \citep{2016arXiv160802013S}. In the future, when light-cone synthetic catalogs are included in the framework, we will incorporate a broader range of SDSS galaxies, as well as objects with deeper imaging, e.g., from CFHTLS \citep{Hudelot12} or DES \citep{2005astro.ph.10346T}, and spectroscopy, e.g., from GAMA \citep{2011MNRAS.413..971D}, DEEP2 \citep{Newman13}, \rev{and DESI \citep{2016arXiv161100036D}}.

In order to compare SDSS colors with synthetic galaxy colors, we use the SDSS apparent magnitudes to construct $K$-corrected absolute magnitudes. First, we select SDSS galaxies in the redshift range $0.06<z<0.09$, where the variation of the distribution of $K$-corrected colors with redshift is small, and correct for Galactic extinction (we implicitly assume that the color evolution for $0<z<0.09$ is negligible when comparing to the single-epoch catalogs). We then use the \textsc{kcorrect} code of \citet{blanton2007} to find the rest-frame spectral energy distributions (SEDs) and obtain $K$-corrected absolute magnitudes for each SDSS galaxy (e.g., $M_i$ for $i$-band absolute magnitude).
{Since different catalogs include galaxy colors that are $K$-corrected to different redshifts and this difference cannot be eliminated by the reader interface, this test $K$-corrects the SDSS data to the same redshift that each of the catalogs uses for its passbands.}

To minimize the effects of incompleteness {in the validation data set}, we construct a volume-limited sample by applying a cut in $r$-band absolute magnitude $M_r<M_{r,\text{max}}$, where $M_{r,\text{max}}$ is chosen to be the value of the 85th percentile of the SDSS $M_r$ distribution in a narrow redshift bin at $0.089<z<0.09$. The same $M_r<M_{r,\text{max}}$ cut is also applied to the synthetic catalogs. We then compare the $K$-corrected colors of the volume-limited samples from SDSS and from the synthetic galaxy sample. 

To obtain a quantitative estimate of the level of difference between the SDSS and synthetic color distributions, we calculate the two-sample Cram\'{e}r-von Mises (CvM) statistic \citep{anderson62}. The CvM test is a nonparametric test for whether multiple data sets are drawn from the same probability distribution, similar to the Kolmogorov--Smirnov (K--S) test. However, instead of only looking at the maximal difference in the cumulative distribution function (CDF), as is done in the K--S test, the CvM test statistic $\omega$ (defined below), calculates the average L$^2$ distance across the entire CDF.  As a result, it is more sensitive to differences in the tails of the distribution than the K--S statistic, and constrains the closeness of two CDFs at every point along them.  

\begin{figure}[tb!]
\centering\includegraphics[width=\columnwidth]{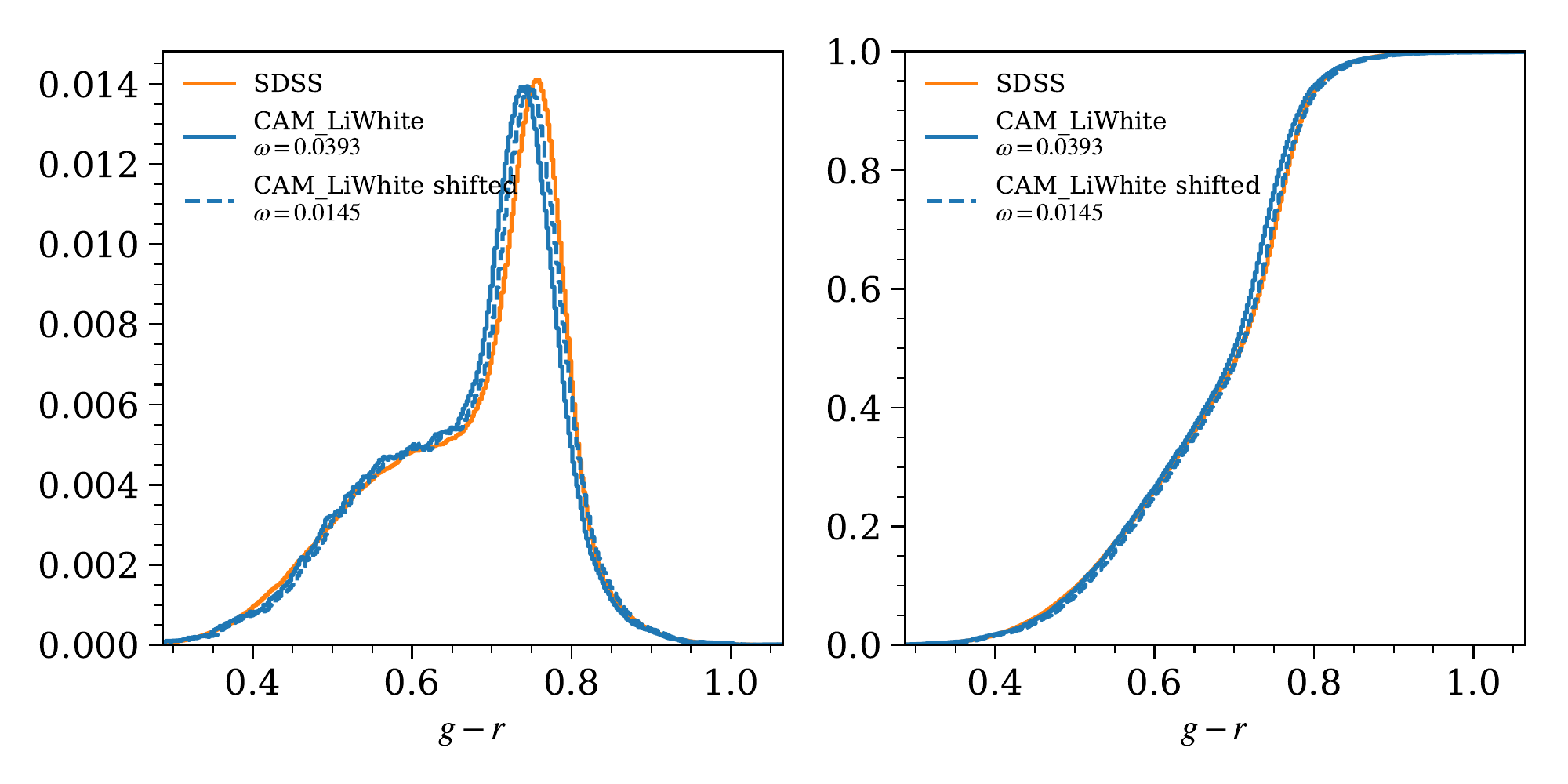}
\caption{\label{fig:color_cdf_sdss}%
Comparison of the color distributions of SDSS galaxies and of the synthetic colors from the CAM\_LiWhite catalog. The left panel shows the probability density functions and the right panel shows the cumulative distribution functions for the $g-r$ colors in each sample. The orange and blue solid curves are the distribution of SDSS colors and synthetic colors, respectively. The dashed blue curve is the distribution of the synthetic color after applying a constant shift to match its median color value to that of SDSS. This should remove the effect of zero-point offsets if they are the main contributors to differences in the color distribution. The definition of $\omega$ can be found in \autoref{eq:color_omega}.}
\end{figure}

The CvM statistic is calculated from the formula
\begin{equation}
\label{eq:color_omega}
\omega^2 = \int_{-\infty}^{+\infty}\big(F_1(x) - F_2(x)\big)^2\mathrm{d} H(x)
\end{equation}
where $F_1(x)$ and $F_2(x)$ are the CDFs of each sample estimated from the data, $H(x) = (n_1F_1(x) + n_2F_2(x))/N$, and $n_1$ and $n_2$ are the numbers of objects in the two samples, with $N=n_1+n_2$. In our case, $F_1(x)$ and $F_2(x)$ are the CDFs for one SDSS color (e.g., $g-r$) and the equivalent synthetic color; $\omega$ provides a measure of the RMS difference between these two CDFs.  As an example, \autoref{fig:color_cdf_sdss} shows the CDFs of the color distribution from SDSS and one of the synthetic catalogs. To remove potential zero-point offsets between SDSS and the synthetic catalogs (whether due to photometric zero-point uncertainties or issues with $K$-corrections), we apply a constant shift to the synthetic galaxy colors so that their median matches the median SDSS color;  we calculate $\omega$ for both unshifted and shifted colors. {The current criterion for} a synthetic catalog to pass the color test is that the $\omega$ calculated from the shifted colors must be smaller than 0.05 for all four SDSS colors ($u-g$, $g-r$, $r-i$, and $i-z$), i.e., the RMS difference between the color CDFs must each be smaller than this threshold.  {This criterion is very stringent and may not reflect the final actual requirements,} but we do expect that LSST will have stringent requirements on the distribution and evolution of galaxy colors to mitigate systematic errors in photometric redshifts.

\begin{figure}[tb!]
\centering\includegraphics[width=\columnwidth]{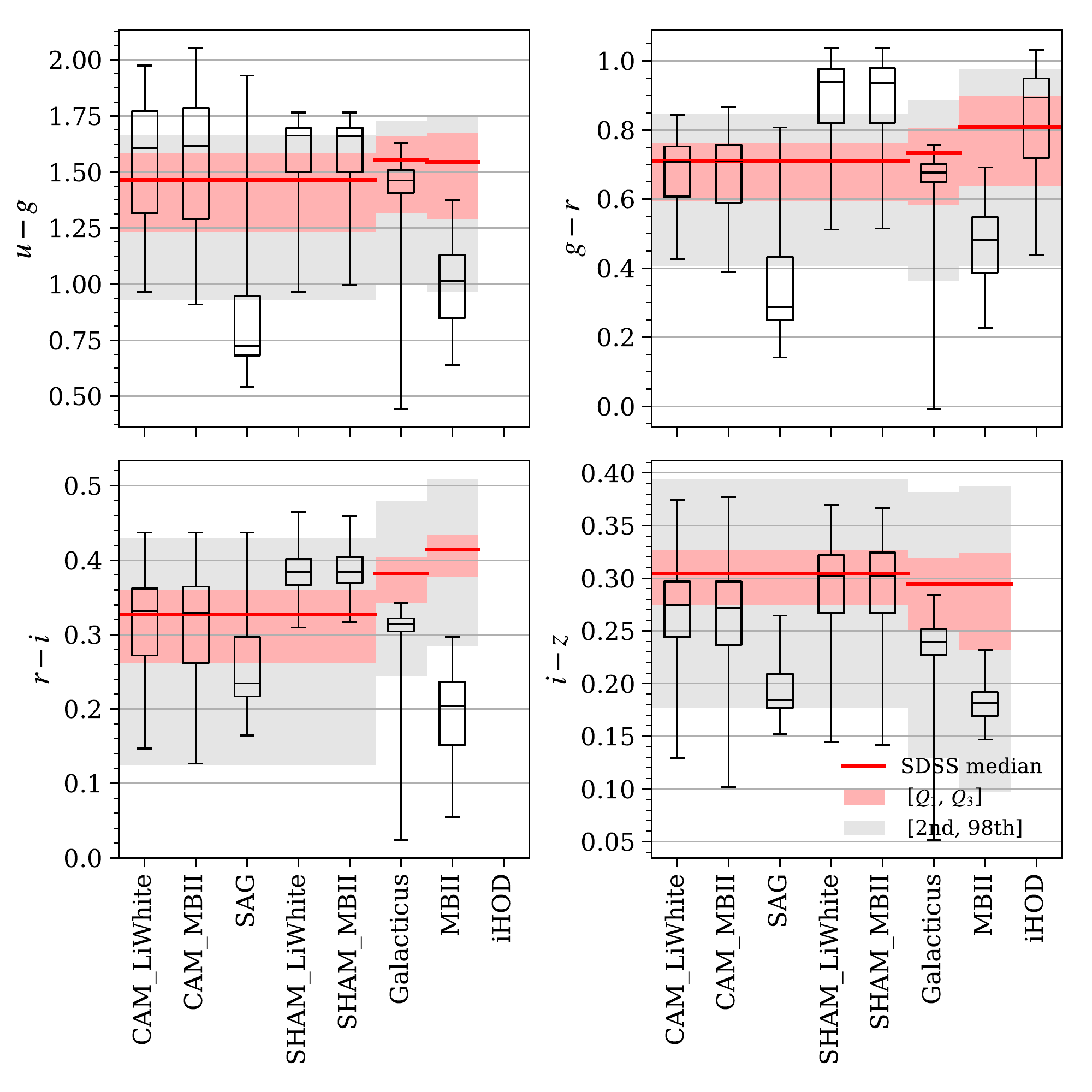}
\caption{\label{fig:color_sdss}%
Summary plot illustrating the differences in the color distributions for those synthetic catalogs that provide color information. The results are shown in the form of box plots for the catalogs and bands for the validation data set (SDSS). For the  validation data set, the red line marks the median, the red band shows the interquartile range, and the gray band shows the range between the 2.5$^\text{th}$ and 97.5$^\text{th}$ percentiles. The synthetic catalogs are represented by the box plots, with the boxes showing the median and the interquartile range and the ends of the extended lines showing the 2.5$^\text{th}$ and 97.5$^\text{th}$ percentiles. The SDSS color distributions vary because for each synthetic catalog, the SDSS colors are $K$-corrected to the redshift corresponding to a given catalog's snapshot. The iHOD catalog only provides the $g-r$ color. }
\end{figure}

A set of summary plots for all catalogs with available colors is shown in \autoref{fig:color_sdss}.
The SHAM catalogs agree well with SDSS in $i-z$ but are redder in $u-g$, $g-r$ and $r-i$. The CAM catalog agrees well with SDSS in $g-r$ and $r-i$, but not as well in $u-g$ and $i-z$, although the difference in $i-z$ might be due to a zero-point offset. Different releases of SDSS, and moderately different redshift ranges, were used in the production of the abundance-matching-based catalogs and the validation catalog, and these two factors likely contribute to the differences. The two stellar mass functions (from MBII and from \citealt{LiWhi09}) yield only a very small difference in the color distributions. The two catalogs from semi-analytic models do not produce very realistic color distributions, with SAG color distributions exhibiting a stronger bimodality than SDSS and Galacticus having long blue tails. 
{We see that some of the features in the color distributions are not captured by the summary statistics. Although the framework requires each test to report a summary statistic for the ease of quick overall comparison, for tests like the color distribution test, the figures are essential components as they provide more information for the catalog users.}

{
\subsection{How These Case Studies Influence Our Design}
\label{sec:findings}

As mentioned above, our case studies have helped us identify several features that are particularly useful for a validation framework like DESCQA. Here we summarize these features. 

\begin{enumerate}
\item \textit{Uniform interfaces for both reading catalogs and executing tests.} To carry out the validation tests presented in our case studies, we can easily see the necessity of the two main components of the DESCQA framework: the reader and test interfaces. By providing a uniform interface, it enables the users to access different catalogs in a uniform way and also standardizes the necessary elements of a validation test. 

\item \textit{Allowing absent quantities.} Since each validation test only accesses a subset of quantities, the framework should not impose a global set of required quantities for all catalogs. Each catalog will be validated by the tests for which it contains the quantities needed. Thus, the reader interface should provide a method for checking available quantities.

\item \textit{Documenting the intrinsic differences in quantity definitions.} Some quantities may be defined differently (e.g., halo mass, magnitudes) in different catalogs. In the cases where these differences cannot be homogenized by the reader interface, the difference should be recorded in the metadata, and be exposed to the tests through the reader interface. It is up to each test developer to decide how to deal with these intrinsic differences.

\item \textit{Adaptive tests.} In the cases where intrinsic differences that cannot be homogenized exist among the catalogs, sometimes it is computationally more efficient for the test to change its configuration on the fly to adapt to each catalog. For example, in our case studies, the HMF validation data set is computed analytically and hence can be tuned to different redshifts to match the catalogs. Similarly, the color distribution test also applies $K$-corrections to the validation data set to match the specification in the catalogs.

\item \textit{Configurable tests.} For a given test, it may be desirable to have some variants that use different validation data sets, passing criteria, or validation ranges. Hence, the validation test interface should provide a convenient method to allow quick changes of these settings. In practice, these settings can be specified in a configuration file that is read in by the test interface. 

\item \textit{Providing both numerical and visual results.} A potential concern with a highly automated validation system is that some important issue is buried under a simple pass-or-fail result. Hence, we encourage the validation test developers to create plots in their test implementation, and we have designed an automated framework to manage such plots and display them on the Web interface. The framework also collects a numerical score from the tests to present in the summary view (\autoref{fig:web}) so that users can spot any potential issues more quickly.

\end{enumerate}
}


\section{Conclusion and Outlook}
\label{sec:conclusion}

In this paper, we have presented DESCQA, a framework that enables the automated validation of synthetic sky catalogs. 
The major aims of the framework are (1) to provide simulators an easy interface to test the quality of their catalogs and (2) to provide the LSST~DESC and larger community a platform that enables them to easily choose synthetic catalogs that best fulfill their needs to test their analysis pipelines.
The necessity of this framework arises from the fact that, with the wide variety of cosmological investigations possible with LSST---weak to strong lensing, cluster cosmology, LSS measurements, and supernova distances---no one single synthetic catalog will be optimal for every task. {For example,  obtaining large volumes for LSS investigations will clearly only be possible with limited mass resolution, while on the other hand, photo-$z$ tests do not necessarily require large volumes, but rather excellent modeling of the color distribution. 
Hence, a framework that can test a wide range of synthetic sky catalogs against a large set of different target requirements is essential in systematically preparing for LSST science.}

{
The goal of the DESCQA framework is to minimize the burden on both the catalog creators and users when they deal with an inhomogeneous set of catalogs and tests. 
We have designed common APIs for both accessing and testing the synthetic sky catalogs with the Python Programming Language, and have also built a Web interface for ease of comparison. Nevertheless, the fundamental challenge here is to design a framework that can actually respond to the scientific needs of catalog validation. 
To meet this challenge during the development of the framework, we have selected a set of realistic synthetic sky catalogs and validation tests to test and improve our framework.

Although these interim synthetic sky catalogs and validation tests are not necessarily the final product or requirements that LSST~DESC will eventually adopt, they have provided useful insights into questions such as how to homogenize a diverse set of synthetic sky catalogs and how to design meaningful validation tests, as we have summarized in this paper. 
Thanks to the use of these realistic trial catalogs and tests during our development process, we have already identified several needed improvements for upcoming LSST~DESC Data Challenges.
For example, although the code itself is maintained in a GitHub repository$^\text{\ref{fn:github}}$ and all the outputs are stored on the NERSC filesystem, a more rigorous catalog and test version control system for the framework is still needed. We also need to improve the ability of the framework to process even larger sky catalogs efficiently and to enable a convenient way to download catalogs of interest (currently only available to LSST~DESC members in our NERSC LSST project space).

Another major step is to include light-cone catalogs, which is essential for realistic comparison with photometric data.
In addition, the set of validation tests will also be considerably extended to cover a large range of possible LSST~DESC projects. During our development process, it became clear that more consideration is needed when designing the catalog requirements. In particular, the validation tests need to carefully handle the intrinsic differences between catalogs that cannot be homogenized by the framework; we have highlighted many issues of this kind here, {as summarized in \autoref{sec:findings}}. 
With this study and the implementation of the DESCQA framework, we have made an important step toward the full utilization of the wide variety of synthetic sky catalogs.
}

\acknowledgments
This paper has undergone internal review by the LSST Dark Energy Science Collaboration. The internal reviewers were Douglas Applegate, Deborah Bard, and Dominique Boutigny.  

The contributions from the authors are listed below.
Y.-Y.M.\ is one of the main developers of the DESCQA framework, led the transition from the original FlashTest framework to the current framework, generated the MBII DMO halo catalogs and merger trees, provided the SHAM-ADDSEDS galaxy catalogs, developed the validation tests, and contributed to the manuscript.
E.K.\ is one of the main developers of the DESCQA framework, provided the Galacticus catalogs, developed the validation tests, and contributed to the manuscript.
K.H.\ initiated this project, has overseen the design and development as a co-convener of the Cosmological Simulation Working Group, and contributed substantially to the manuscript.
T.D.U.\ contributed to the current framework and also the transition from the original FlashTest framework to the current framework, and assisted in integrating catalogs and tests to reach the current validation capability.
A.J.B.\ is the main developer of the Galacticus model, assisted in generating the Galacticus galaxy catalog, and contributed to the manuscript.
D.C.\ provided the CAM catalogs and contributed to the DESCQA framework and the manuscript.
S.A.C.\ worked on the SAG catalog development.
J.D.\ was the main developer of the SHAM-ADDSEDS catalogs.
T.D.M.\ helped initiate the project and contributed to the MBII simulations and catalogs.
S.H.\ contributed to the manuscript and provided advice on the overall framework functionality and error metrics.
A.P.H.\ contributed code for generating the CAM catalogs and oversaw D.C.'s implementation of the method into the DESCQA framework.
J.B.K.\ contributed to the initial implementation of the framework and initial color tests.
K.S.K.\ participated in this project as a co-convener of the Cosmological Simulation Working Group.
F.L.\ provided the reader interface to the MBII hydrodynamic galaxy catalog.
Z.L.\ produced the halo mass function test, and contributed to the project discussions and manuscript.
R.M.\ helped initiate the project and contributed to the iHOD catalog and the manuscript.
J.A.N.\ provided advice on metrics and methods, especially for the galaxy color distribution test.
N.P.\ and E.P.\ provided the SAG galaxy catalog and contributed to the manuscript.
A.P.\ helped coordinate infrastructure to produce the Galacticus catalogs.
P.M.R.\ created the original prototype for the FlashTest framework, established the mapping from the FlashTest concepts to the DESCQA ones, and contributed to several utilities in the framework.
A.N.R.\ contributed to the SAG catalog.
A.T.\ assisted in generating the MBII hydrodynamic galaxy catalog, provided validation data sets used in several tests, and contributed to the manuscript.
C.V.-M.\ contributed to the SAG catalog generation.
R.H.W.\ helped initiate the project and contributed to the manuscript, overall project guidance, and to the SHAM-ADDSEDS catalogs.
R.Z.\ developed the galaxy color distribution test and contributed to the manuscript.
Y.Z.\ provided the iHOD galaxy catalog and contributed to the manuscript.

The MBII hydrodynamical simulation was run on the BlueWaters facility at the National Center for Supercomputing Applications.
Argonne National Laboratory's work was supported under the DOE contract DE-AC02-06CH11357. 
Part of this work uses the computational resources at the SLAC National Accelerator Laboratory, a U.S.\ DOE Office; we thank the support of the SLAC computational team.
The framework used for this project is derived from a framework that was developed for FLASH and applied to the Dark Energy Survey by the DOE-supported ASC/Alliances Center for Astrophysical Thermonuclear Flashes at the University of Chicago; we thank Gus Evrard, Michael Busha, and Andrey Kravtsov for their contributions. We thank Joanne Cohn for discussions on validation tests.
This research has made use of NASA's Astrophysics Data System.

Y.-Y.M.\ is supported by the Samuel P.\ Langley PITT PACC Postdoctoral Fellowship and was supported by the Weiland Family Stanford Graduate Fellowship.
S.A.C.\ acknowledges grants from Consejo Nacional de Investigaciones Cient\'{i}ficas y T\'{e}cnicas (PIP-112-201301-00387), Agencia Nacional de Promoci\'{o}n Cient\'{i}fica y Tecnol\'{o}gica (PICT-2013-0317), and Universidad Nacional de La Plata (UNLP 11-G124), Argentina.
T.D.M.\ acknowledges funding from NSF ACI-1614853, NSF AST-1517593, NSF AST-1009781, NSF AST-1616168, and the BlueWaters PAID program.
F.L.\ and R.M.\ acknowledge the support of the Department of Energy Early Career Award program.
Z.L.'s work was partially funded by the Scientific Discovery through Advanced Computing (SciDAC) program funded by the U.S.\ DOE Office of Advanced Scientific Computing Research and the Office of High Energy Physics.
J.A.N.\ and R.Z.\ acknowledge support from DOE under grant DE-SC0007914. 
N.P.\ acknowledges support from Fondecyt 1150300, BASAL PFB-06 Centro de Astrof\'\i sica y Tecnolog\'\i as Afines.
P.M.R.\ acknowledges the support and hospitality of the University of Michigan Department of Physics, where initial framework development was done during a sabbatical visit.
C.V.-M.\ was supported by a fellowship from CONICET, Argentina. 
Y.Z.\ acknowledges the support of the CCAPP Fellowship.

The DESC acknowledges ongoing support from the Institut National de Physique Nucl\'eaire et de Physique des Particules in France; the Science \& Technology Facilities Council in the United Kingdom; and the Department of Energy, the National Science Foundation, and the LSST Corporation in the United States.  DESC uses the resources of the IN2P3 Computing Center (CC-IN2P3--Lyon/Villeurbanne - France) funded by the Centre National de la Recherche Scientifique; the National Energy Research Scientific Computing Center, a DOE Office of Science User Facility supported by the Office of Science of the U.S.\ Department of Energy under contract No.\ DE-AC02-05CH11231; STFC DiRAC HPC Facilities, funded by UK BIS National E-infrastructure capital grants; and the UK particle physics grid, supported by the GridPP Collaboration.  This work was performed in part under DOE contract DE-AC02-76SF00515.

\software{\rev{Python~2.7, Python~3.6,}
Astropy \citep{astropy},
h5py (\href{http://www.h5py.org}{h5py.org}),
Matplotlib \citep{matplotlib},
NumPy \citep{numpy},
SciPy \citep{scipy}
}

\appendix

\section{Simulations and Synthetic Sky Catalogs}
\label{sec:catalogs}

{Here we describe the set of synthetic catalogs used in our case studies (\autoref{sec:validation}).}
We first provide a description of the MassiveBlack-II simulations (\autoref{sec:mbII}), for both the hydrodynamical and gravity-only runs. In the following subsections, we discuss the six different methods (one hydrodynamical simulation, one HOD-based model, two SHAM-based models, and two SAMs) to generate the eight synthetic catalogs used in \autoref{sec:validation}.
{All of the synthetic methods used to generate the catalogs are applied to the same dark matter structures (i.e., halos and merger trees) of the MBII DMO simulation. A brief summary of these catalogs is listed in \autoref{tab:models}.}

\subsection{The MassiveBlack-II (MBII) Simulations}
\label{sec:mbII}

MBII is a state-of-the-art, high-resolution
cosmological hydrodynamic simulation \citep{Khandai15} of structure formation with subgrid model physics described in detail below. A companion simulation, MBII DMO, uses the same volume, resolution, cosmological parameters, and initial conditions but only takes gravitational forces into account \citep{2015MNRAS.453..469T}. Both of these simulations have
been performed in a cubic periodic box of size $142.45 \, \text{Mpc}$ on a side
using the cosmological TreePM Smooth
Particle Hydrodynamics (SPH) code \textsc{p-gadget}, which is a hybrid version of the parallel code,
\textsc{gadget2} \citep{Springel05}, that has been upgraded to
run on petascale supercomputers. The total number of dark matter
particles in both simulations is $1792^3$ with an equal (initial) number of gas particles in the hydrodynamical simulation. 

\begin{table}[htb!]
\caption{\label{tab:param}Simulation parameters: Box size ($L_\text{box}$), force softening length ($\epsilon$), number of particles ($N_\text{part}$), mass of dark matter particle ($m_\text{DM}$), and mass of gas particle ($m_\text{gas}$).}
\begin{center}
\begin{tabular}{c|c|c}
Parameters & Hydrodynamical & Dark Matter-Only\\ 
\hline  
$L_\text{box}$ (Mpc)& 142.45 & 142.45\\ 
$\epsilon$ (kpc)& 2.64 & 2.64\\
$N_\text{part}$ & $2 \times 1792^{3}$ & $1792^{3}$\\
$m_\text{DM}$ ($\msun$) & $1.6 \times 10^{7}$ & $1.9 \times 10^{7}$\\
$m_\text{gas}$ ($\msun$) & $3.1 \times 10^{6}$ & --\\
\end{tabular}
\end{center}
\end{table}

The cosmological parameters are chosen for consistency with WMAP7
\citep{Komatsu11}, with amplitude of matter
fluctuations set by $\sigma_{8} = 0.816$, the scalar spectral index $n_{s} = 0.96$,
matter density parameter $\Omega_{m} = 0.275$, cosmological constant
density parameter $\Omega_{\Lambda} = 0.725$, baryon density parameter
$\Omega_{b} = 0.046$ (in MBII), and Hubble parameter $h = 0.702$. 
\autoref{tab:param} lists the box size ($L_\text{box}$), force
softening length ($\epsilon$), total number of particles including
dark matter and gas ($N_\text{part}$), mass of dark matter particles
($m_\text{DM}$), and mass of gas particles ($m_\text{gas}$) for
the two simulations. The major results from the hydrodynamical simulation, MBII, are available in \cite{Khandai15}. In addition to gravity and SPH, MBII also includes the physics of a multiphase interstellar medium model with star formation \citep{Springel03}, and black hole accretion and feedback \citep{Springel05,DiMatteo12}.
Radiative cooling and heating processes are included \citep[as
in][]{Katz96}, as is photoheating due to an imposed homogeneous
ionizing ultraviolet background.

For the analysis of the MBII DMO (gravity-only) simulation, we use \textsc{Rockstar}\footnote{\https{bitbucket.org/gfcstanford/rockstar} (commit \#ca79e51)}, a six-dimension phase-space halo finder, to identify halos and subhalos \citep{2013ApJ...762..109B}, and use \textsc{Consistent Trees}\footnote{\https{bitbucket.org/pbehroozi/consistent-trees} (commit \#2ddc70a)} to build the halo merger trees \citep{2013ApJ...763...18B}.
The halo catalogs and merger trees are available on the NERSC filesystem and will be made publicly available.

The halos are defined with spherical overdensity at virialization \citep{Bryan1998}. At $z=0$, this virial overdensity ($\Delta_\text{vir}$) is approximately $97.7$ for this cosmology.
Subhalos are defined as halos whose centers are within the virial radius of any other larger halo.
When building the merger trees, we skip some very close-timed snapshots and use in total $177$ snapshots from the simulation, with the earliest snapshot at $z=20$. In the halo catalogs that we provide for the synthetic catalog creators, each halo or subhalo in the catalog must have at least 20 particles associated with it. The catalog creators may apply more stringent halo mass cuts if required.

\subsection{MBII Galaxy Catalog}
\label{sec:mbII-galaxy}

The MBII hydrodynamical simulation was analyzed by applying an FOF procedure to dark matter particles, with a dimensionless linking length of $b=0.2$. 
Gas, star, and black hole particles were then associated to their nearest dark matter particles.
Subhalos were identified with the subhalo finder SUBFIND~\citep{Springel01}. The galaxy stellar mass is the total mass of all the star particles bound to the subhalo. The SED of star particles in MBII are generated using the \textsc{Pegase.2} stellar population synthesis code \citep{1997A&A...326..950F,1999astro.ph.12179F}, based on the ages, masses, and metallicities of the stars, with the assumption of a Salpeter initial mass function. Nebula (continuum and line) emissions are also added to each star particle SED, along with a correction for absorption in the intergalactic medium using the standard \cite{1996MNRAS.283.1388M} prescription. The SED of a galaxy is then obtained by summing the SEDs of all the star particles in the galaxy, from which SDSS-band luminosities are calculated, based on the respective filter.  More details can be found in~\cite{Khandai15}. 

Here we use the SUBFIND halo catalogs for the MBII hydrodynamical simulation as those match the published version, but we use the \textsc{Rockstar--Consistent Trees} catalogs for the MBII DMO run as they provide more robust merger histories \citep{2014MNRAS.441.3488A}. 
Therefore, we caution the reader that when comparing results in \autoref{sec:validation}, it should be kept in mind that some differences in the tests that rely on halo masses are to be expected because the MBII hydrodynamic and DMO runs use different halo finders and different mass definitions.


\subsection{Improved Halo Occupation Distribution (\ihod{}) Model}
\label{sec:ihod}

The \ihod{} model~\citep{zm15a,zm15b,zm17} aims to provide a \emph{probabilistic} mapping between halos and galaxies, assuming that the
enormous diversity in the individual galaxy assembly histories inside similar halos would reduce to a
stochastic scatter about the \emph{mean} galaxy-to-halo connection by virtue of the central limit theorem.
Therefore, the key is to derive the conditional probability distribution of host halos at fixed galaxy properties, $ P(\mathbf{\tilde{h}}\mid\mathbf{\tilde{g}})$, 
where $\mathbf{\tilde{g}}$ and $\mathbf{\tilde{h}}$ are the corresponding vectors that describe the most
important sets of properties.
For $\mathbf{\tilde{g}}$, those properties can be the stellar mass, central/satellite dichotomy, color, velocity, and alignment, and for $\mathbf{\tilde{h}}$ the dark matter mass, concentration, and tidal environment.

Building on the global HOD parameterization of \citet{leauthaud11}, \citet{zm15a} developed the \ihod{} formalism to solve the mapping between galaxy stellar mass and halo mass, i.e., $P(M_h | M_*)$, using the
spatial clustering and the galaxy--galaxy lensing of galaxies in SDSS.
Compared to the traditional HOD methods, \ihod{} can include ${\sim}84\%$ more galaxies while taking into
account the stellar mass incompleteness of galaxy samples in a self-consistent fashion.

In order to link galaxy colors to the underlying dark matter halos and constrain galaxy quenching,
\citet{zm15b} extended the \ihod{} model to describe galaxies of different $g-r$ colors, i.e.,
$\mathbf{\tilde{g}}\equiv\{M_*,\,g-r\}$, by considering two popular quenching scenarios: (1) a ``halo''
quenching model in which halo mass is the sole driver for turning off star formation in both central and
satellite galaxies and (2) a ``hybrid'' quenching model in which the quenched fraction of galaxies
depends on their stellar mass while the satellite quenching has an extra dependence on halo mass.
\citet{zm15b} found that the halo quenching model provides significantly better fits to the clustering and galaxy--galaxy
lensing of blue galaxies above stellar masses of $10^{11}\,\msun$. The best-fitting \ihod{} quenching model of $P(M_h | M_*,\,g-r)$ also correctly predicts the average halo mass of the red and blue centrals, showing excellent
agreement with the direct weak-lensing measurements of central galaxies~\citep{mandelbaum15}.  The \ihod{}
modeling of galaxy colors provides strong evidence that the physical mechanism that quenches star formation
in galaxies above stellar masses of $10^{10}\,\msun$ is tied principally to the masses of their dark matter halos rather than
the properties of their stellar components or halo age.

\citet{zm17} further demonstrated that the \ihod{} model provides an excellent description of the environmental dependence and conformity of galaxy colors observed in SDSS. The current \ihod{} model, as constrained by the clustering and galaxy--galaxy lensing of red and blue galaxies in SDSS,
also correctly reproduces the stellar mass functions within each color observed by SDSS. For the purpose of
this paper, we populate the halo catalog using the best-fit parameters from \citet{zm17}.

\subsection{SHAM-ADDSEDS Model}
\label{sec:AMM}
This synthetic catalog is a combination of the SHAM \citep[see e.g.,][]{Kravtsov2004,Vale2004,Vale2006,Conroy2006}  and the \textsc{ADDSEDS} algorithm (R.~H.~Wechsler et al.\ 2018, in preparation; J.~DeRose et al.\ 2018, in preparation) which is explained in more detail below.
The SHAM technique is a generic scheme to connect one galaxy property (e.g., stellar mass or luminosity) with one (sub)halo property (e.g., virial mass) by assuming an approximately monotonic relation between these two properties.
The two properties are matched at the same cumulative number density, and the resulting galaxy catalog, by explicit construction, preserves the input stellar mass (or luminosity) function. 

Common choices of the matching (sub)halo properties include $M_\text{peak}$ and $V_\text{peak}$, which are the mass and the maximal circular velocity, respectively, at their peak values along the accretion history of the (sub)halo. 
Scatter between the galaxy and (sub)halo properties can be introduced into the matching procedure. For a constant log-normal scatter in stellar mass or luminosity, one can follow the procedure in \citet{Behroozi10}: first deconvolve the scatter from the stellar mass (or luminosity) function, match the two properties, and finally add a random log-normal scatter in the catalog.

To generate the synthetic catalogs used in this work, we use a publicly available SHAM code,\footnote{\https{bitbucket.org/yymao/abundancematching}} which follows the procedure we outlined above, to match the (sub)halo $V_\text{peak}$ function to the stellar mass functions from \cite{LiWhi09} and MassiveBlack-II, respectively. 
In both cases, we adopt a constant log-normal scatter of 0.15\,dex in stellar mass \citep[see, e.g.,][]{Reddick2013,Gu2016,Lehmann2017}.
Once we obtain the stellar mass for each synthetic galaxy, we also assign an absolute $r$-band magnitude by simply matching the stellar mass function to the luminosity function of \cite{Bernardi2013}. 

We then further generate multiband magnitudes using the ADDSEDS algorithm. For each synthetic galaxy, we measure the projected distance to its fifth nearest neighbor. We then bin galaxies in absolute $r$-band magnitude and rank-order them in terms of this projected distance.
We compile a training set consisting of the magnitude-limited spectroscopic SDSS DR6 VAGC cut to $z<0.2$ and local density measurements from \cite{cooper2006}. This training set is rank-ordered the same way as the simulation. Each simulated galaxy is assigned the SED from the galaxy in the training set with the closest density rank in the same absolute magnitude bin. The SED is represented as a sum of templates from \cite{blanton2007}, which can then be used to shift the SED to the correct reference frame and generate magnitudes in SDSS bandpasses. In our case, we assume that all of our galaxies are at redshift $z=0$, and the magnitudes are $K$-corrected to $z=0$.

\subsection{Conditional Abundance Matching (CAM) Model}
\label{sec:CAMM}
This synthetic galaxy catalog is created using the CAM technique \citep{Hearin14}. We provide a brief description of the catalog and technique here, and point the reader to \citet{1705.06347} for details.  The CAM technique is similar to the SHAM technique (see \autoref{sec:AMM}), but further assigns a secondary galaxy property (e.g., specific star formation rate) according to a secondary (sub)halo property (e.g., mass accretion history). In this work, the primary galaxy property is the stellar mass, which is assigned to the $V_\text{peak}$ of (sub)halos using a simple SHAM technique.
We include a fixed log-normal scatter $\sigma_\text{SMHM}$ in stellar mass at fixed $V_\text{peak}$.
We deconvolve the scatter from the stellar mass function before matching.

As for the secondary properties, specific star formation rates (sSFRs) are assigned to galaxies such that there is a correlation between sSFR and $a_{1/2}$ at fixed stellar mass, where $a_{1/2}$ is the scale factor at which the (sub)halo reached half its peak mass.  This step requires drawing from a $P(\text{sSFR}|M_*)$ distribution, here based on the Main Galaxy Sample from the SDSS Data Release 7 \citep[DR7;][]{Padmanabhan2008,Abazajian2009}, specifically a re-reduction of DR7 in the form of the NYU-VAGC LSS sample \citep{2005AJ....129.2562B}. sSFRs are taken from the MPA-JHU DR7 catalog based on the method of \cite{Kauffmann03}.  The strength of this correlation is encoded in a rank-order correlation statistic, $\rho_\text{sSFR}$, which can generally take values in the range $[-1,1]$, perfect anticorrelation to perfect correlation.  The result is that our model has two explicit parameters that can be tuned:  $\sigma_\text{SMHM}$ and $\rho_\text{sSFR}$.  For this study, we use fiducial values of 0.15 and 1.0, respectively.
Absolute magnitudes in five bands ($ugriz$), $K$-corrected to $z=0$, are associated with each galaxy in the synthetic catalog by selecting a galaxy in the NYU-VAGC LSS catalog with similar stellar mass and sSFR, and carrying over each of the magnitudes.  In this way, the conditional stellar mass, sSFR, and colors are preserved in the synthetic catalog.

\subsection{Semi-analytic Galaxies (SAG) Model}
\label{sec:SAG}

The Semi-analytic Galaxies (SAG) approach is based on the model developed by \citet{Springel01}, which, as is usual with semi-analytic models, combines merger trees extracted from a gravity-only cosmological simulation with a set of coupled differential equations for the baryonic processes taking place within these merger trees as time evolves.
The most up-to-date version of the SAG model has been further improved from the model described in \citet{Gargiulo15}; however, those improvements are not used in the current study in order to achieve good performance.

The model used here assumes that the hot gas in dark matter halos is isothermally distributed, with an initial mass calculated using the cosmic baryon fraction.
This hot gas cools to form an exponential disk where stars form quiescently.
Gas cooling takes place only in central galaxies (i.e., the galaxy residing in the main subhalo of a given dark matter halo); the hot gas atmosphere is stripped instantly when a galaxy becomes a satellite (strangulation scheme).
Stars also form through starbursts, which can be triggered by mergers and disk instabilities contributing to bulge formation.
In that case, the gas is transferred to a reservoir that is continuously consumed by star formation in a given timescale.
This reservoir can be modified by successive mergers and instabilities \citep{Gargiulo15}.
Bursts are the main channel for supermassive black hole growth.
Gas accretion onto these objects produces AGN feedback \citep{Lagos08}.
The stars formed in each star formation event produce a number of supernovae depending on the selected initial mass function. These supernovae reheat the cold gas transferring it back to the hot phase (supernovae feedback).
Chemical elements produced by stellar winds and supernova explosions (both core-collapse and Type Ia supernovae)
are tracked in different baryonic components, taking into account the lifetime of stars \citep{Cora06}.
The current chemical implementation has been updated with new stellar yields \citep{Gargiulo15}.
Stellar luminosities and colors are modeled using the \citet{2003MNRAS.344.1000B} stellar synthesis models for the stellar populations generated in model galaxies (at each integration time step).

SAG depends on a number of parameters.
These are tuned using the Particle Swarm Optimization technique \citep{Ruiz15}.
For this particular run, we consider a set of best-fitting parameters obtained from the application of SAG to one of the MultiDark gravity-only cosmological simulations \citep[MDPL2;][]{Klypin2016} with a cubic volume of (1475.6\,Mpc)$^3$ and \textit{Planck} cosmological parameters \citep{PlanckCollaboration2013}.
The observational constraints used for this calibration are the stellar mass function and the black hole--bulge mass relation, both at $z=0$.
For the former, we adopt the compilation of data used by \cite{henriques_2015}, while for the latter we combine the data sets from \cite{mcconnell_2013} and \cite{kormendy_2013}.

\subsection{Galacticus}
\label{sec:gal}
Galacticus~\citep{benson_2010b} is another semi-analytic model of galaxy formation that we employ in this paper. Galacticus models the baryonic physics of galaxy growth within an evolving, merging hierarchy of dark matter halos. 
Baryonic processes (including gas cooling and inflow, star formation, stellar and AGN feedback, and galaxy merging) are described by a collection of differential equations that are integrated to a specified tolerance along each branch of the merger tree. Also included are instantaneous transformations, such as starbursts, that are associated with merger events. 
Galacticus is designed to be fully modular, allowing the physical components and processes in galaxies and halos to be interchanged easily. This permits the possibility of running everything from simplistic models based on empirical fitting functions for the rates of key processes through to fully physical models incorporating detailed treatments of chemical enrichment, galaxy and halo dynamics, black hole accretion disks, and feedback.

The output of Galacticus is a catalog of galaxies at all redshifts that includes both physical properties, such as stellar masses, sizes, and morphologies, and observational properties, such as luminosities in any specified bandpass filter. The luminosities are computed by convolving the star-formation history for each galaxy with spectra obtained from stellar population synthesis models. For this paper, we computed rest-frame luminosities in SDSS $ugriz$ filters. 
Note that, for this paper, in order to ensure consistency of the input halo catalogs with the other synthetic methods, we disable a standard convergence-testing feature in Galacticus, which ensures that they reach sufficient temporal and mass resolution.

The parameters of the model are constrained through either particle swarm optimization or Markov Chain Monte Carlo techniques to match a wide variety of data on the galaxy population, including the stellar mass function from $z=0$ to $z=5$, the $z=0$ H\textsc{i} mass function, the galaxy size--mass relation, and the two-point correlation function of galaxies. The resulting models can, in principle, accurately reproduce key observables of the galaxy population across a wide range of redshifts. However, the parameters also depend on simulation details such as the mass resolution, and so, in practice, the parameters need to be tuned for each simulation. For this paper, we used a default parameter set obtained by tuning on Press--Schechter trees, and so we do not expect to find good agreement between the Galacticus catalog and the validation data.

\bibliographystyle{yahapj}
\bibliography{refs,software}

\end{document}